\documentclass[aps,pre,superscriptaddress,reprint,twocolumn,nofootinbib]{revtex4-1}
\usepackage{xspace}
\usepackage{graphicx}
\usepackage{tabularx}
\usepackage{color}
\usepackage{amsmath, amsthm, amssymb}
\usepackage{times}

\newcommand{\ucsd}{Department of Physics, University of California,
                          San Diego, La Jolla CA 92093}
\newcommand{\jhu}{Departments of Physics and Astronomy and Biophysics, Johns Hopkins University, Baltimore MD 21218}

\newcommand{\dri}{\ensuremath{\boldsymbol\delta \vb{r}^i}\xspace}

\newcommand{\rij}{\ensuremath{\hat{\vb{r}}^{ij}}\xspace}
\newcommand{\dij}{\ensuremath{d^{ij}}\xspace}

\newcommand{\eq}{Eq.~}

\newcommand{\fig}{Fig.~}
\newcommand{\vb}[1]{ {\mathbf #1}}
\newcommand{\um}{~\ensuremath{\mu}m\xspace}
\newcommand{\rb}{\ensuremath{\vb{r}}\xspace}
\newcommand{\gv}{\ensuremath{\vb{g}}\xspace}
\newcommand{\gh}{\ensuremath{\hat{\vb{g}}}\xspace}
\newcommand{\ght}{\ensuremath{\hat{\vb{g}}}_T\xspace}
\newcommand{\ghtx}{\ensuremath{\hat{\vb{g}}}_{T,x}\xspace}
\newcommand{\ghx}{\ensuremath{\hat{g}_x}\xspace}
\newcommand{\ghy}{\ensuremath{\hat{g}_y}\xspace}
\newcommand{\like}{\ensuremath{\mathcal{L}}\xspace}

\newcommand{\lam}{\ensuremath{\boldsymbol\Lambda\xspace}}
\newcommand{\lamt}{\ensuremath{\boldsymbol\Lambda_T\xspace}}
\newcommand{\sd}{\ensuremath{\sigma_{\Delta}}}
\newcommand{\sg}{\ensuremath{\sigma_{\mathbf{g}}}}
\newcommand{\sgt}{\ensuremath{\sigma_{\mathbf{g},T}}}
\newcommand{\sgz}{\ensuremath{\sigma_{\mathbf{g},0}}}
\newcommand{\tavg}{\ensuremath{T}}

\newcommand{\Mconf}{\ensuremath{\{M^{i}\}\xspace}}
\newcommand{\davg}[1]{\ensuremath{\left\langle #1 \right\rangle}}
\newcommand{\snr}{\ensuremath{\textrm{SNR}}}
\newcommand{\Ainv}{\ensuremath{\mathcal{A}^{-1}}}

\usepackage{hyperref}
\begin{document}
\title{Cell-to-cell variation sets a tissue-rheology-dependent bound on collective gradient sensing}

\author{Brian A. Camley}
\affiliation{\ucsd}
\affiliation{\jhu}
\author{Wouter-Jan Rappel}
\affiliation{\ucsd}

\begin{abstract}
When a single cell senses a chemical gradient and chemotaxes, stochastic receptor-ligand binding can be a fundamental limit to the cell's accuracy. 
For clusters of cells responding to gradients, however, there is a critical difference: even genetically identical cells have differing responses to chemical signals.
With theory and simulation, we show collective chemotaxis is limited by cell-to-cell variation in signaling. We find that when different cells cooperate the resulting bias can be much larger than the effects of ligand-receptor binding. Specifically, when a strongly-responding cell is at one end of a cell cluster, cluster motion is biased toward that cell. These errors are mitigated if clusters average measurements over times long enough for cells to rearrange. In consequence, fluid clusters are better able to sense gradients: we derive a link between cluster accuracy,  cell-to-cell variation, and the cluster rheology. Because of this connection, increasing the noisiness of individual cell motion can actually {\it increase} the collective accuracy of a cluster by improving fluidity. 
\end{abstract}

\maketitle

Many cells follow signal gradients to survive or perform their functions, including white blood cells finding a wound, cells crossing a developing embryo, and cancerous cells migrating from tumors. Chemotaxis, sensing and responding to chemical gradients, is crucial in all of these examples
\cite{swaney2010eukaryotic,levine2013physics}. Chemotaxis is traditionally studied by exposing single cells to gradients -- but cells often travel in groups, not singly \cite{hakim2017collective,camley2017physical}. Collective cell migration is essential to development and metastasis \cite{friedl2009collective}, and 
can have remarkable effects on chemotaxis. Even when single cells cannot sense a gradient, a cluster of cells may cooperate to sense it.  While collective chemotaxis is our primary focus, this ``emergent'' gradient sensing is found in response to many signals, including soluble chemical gradients (chemotaxis) \cite{theveneau2010collective,malet2015collective,ellison2016cell}, conditioned substrates (haptotaxis) \cite{winklbauer1992cell}, substrate stiffness gradients (durotaxis) \cite{sunyer2016collective} and electrical potential (galvanotaxis) \cite{li2012cadherin,lalli2015collective}. 

Cells can cooperate to sense gradients -- but the physical principles limiting a cluster's sensing accuracy are not settled. For single cells, the fundamental bounds on sensing chemical concentrations and gradients are well-studied \cite{berg1977physics,kaizu2014berg,hu2010physical,hu2011geometry,endres2009maximum,endres2008accuracy,fuller2010external,ueda2007stochastic,andrews2007information,bialek2005physical}, showing unavoidable stochasticity in receptor-ligand binding limits chemotactic accuracy. Is this true for cell clusters? Is a cell cluster simply equivalent to a larger cell? No! There is an essential difference between many clustered cells and a single large cell: even clonal populations of cells can have highly variable responses to signals, due to many factors, including intrinsic variations in regulatory protein concentrations \cite{swain2002intrinsic,niepel2009non,sigal2006variability}. These cell-to-cell variations (CCV) can be persistent over timescales much larger than the typical motility timescale of the cell \cite{sigal2006variability}. CCV has not been addressed in models of collective chemotaxis and it is not
clear whether collective gradient sensing is limited by CCV or by stochastic 
receptor-ligand binding \cite{malet2015collective,camley2016emergent,camley2016collective,varennes2016collective,ellison2016cell,mugler2016limits,cai2016modeling}.

Using a combination of analytics and simulations, we show that unless CCV is tightly controlled, collective guidance of a cluster of cells is limited by these variations: gradient sensing is biased toward cells with intrinsically strong responses. This bias swamps the effects of stochastic ligand-receptor binding. Cell clusters may reduce this error by time-averaging their gradient measurements only if the cells rearrange their positions, creating an unavoidable link between the {\it mechanics} of the cell cluster and its {\it gradient sensing ability}. As a result, surprising new tradeoffs arise: clusters must balance using motility to follow a biased signal with using motility to reduce error, and compromise between reducing noise and increasing cluster fluidity.

\section{Gradient sensing error is dominated by cell-to-cell variation, not receptor noise}

\begin{figure*}[ht!]
 \centering
\includegraphics[width=180mm]{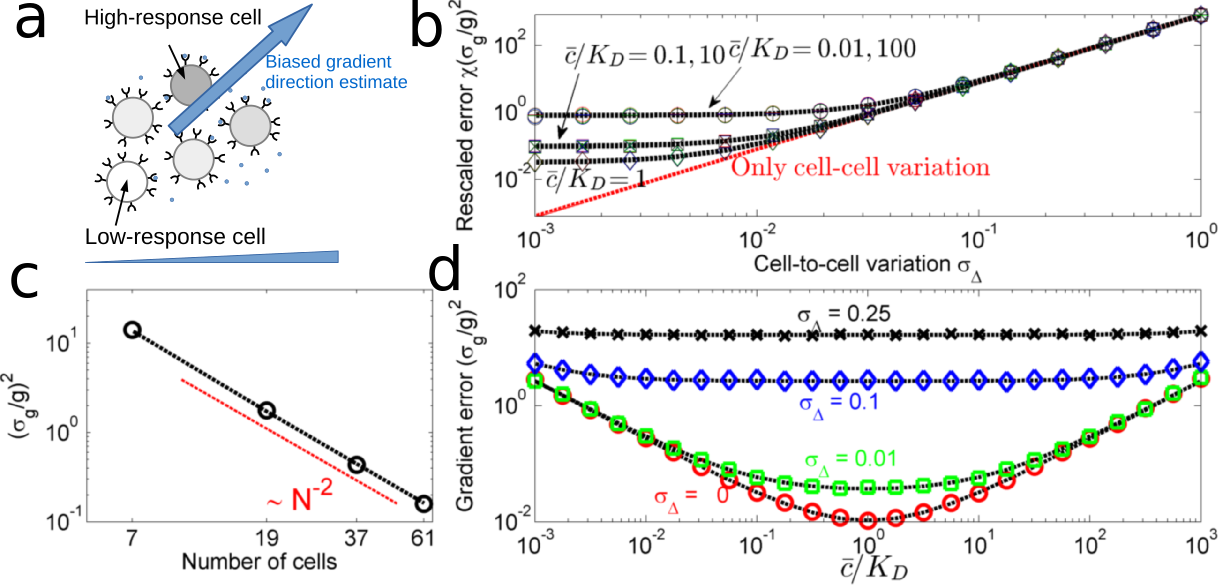}
 \caption{\linespread{1.0}\selectfont{}{\bf Cell-to-cell variation creates systematic biases that can be significantly larger than the effects of receptor-ligand binding}. {\bf a:} Schematic of how cell-to-cell variation can create bias in gradient sensing toward high-signaling and away from low-signaling cells. {\bf b:} Gradient sensing error $\sg^2 = \davg{|\hat{\vb{g}}-\gv|^2}$, derived from numerical maximum likelihood (symbols), is well-approximated by \eq \ref{eq:combined_error} (dashed lines) at low gradient strengths. Symbols are plotted for four cluster sizes: $N = 7, 19, 37,$ and $61$ cells (hexagonally-packed clusters of unit spacing with $Q = 1,2,3,4$ layers, illustrated in Appendix \ref{app:Q}). In all panels, we use $n_r = 10^5$ and $g = 0.05$, in units where the cell-cell spacing is one. {\bf c:} Gradient sensing error decreases as cluster size increases as $\sg^2 \sim N^{-2}$. In this panel, $\sd = 0.23$ and $\bar{c} = K_D$. {\bf d:} Strong CCV can mask concentration-dependence of accuracy. In absence of CCV, gradient sensing accuracy is maximized when $\bar{c} \approx K_D$; this effect is screened when CCV dominates gradient sensing. This panel is shown for $N = 7$ cells.}
 \label{fig:ccv_illus}
 \end{figure*}

We study a two-dimensional  model of gradient sensing with CCV and ligand-receptor dynamics
where cells sense a chemoattractant with concentration gradient $\gv$.  Each cell at position $\rb$ measures local concentration, $c(\rb) = c_0 (1 + \gv \cdot \rb)$,  via ligand-receptor binding, which is stochastic. This noise leads to unavoidable errors in the cluster's estimate of $\gv$. In addition, even if concentration is perfectly sensed, each cell responds differently to a fixed $c$, which models known CCV in signal response \cite{wang2012diverse,samadani2006cellular}. As a result, when the cluster combines measurements from its cells, it may develop a drift in the direction of stronger-responding cells (\fig \ref{fig:ccv_illus}). To combine these effects, we
specify the ``measured'' signal in cell $i$, $M_i$, which is what the cluster believes the chemoattractant signal in cell $i$ to be, including ligand-receptor binding and CCV:
\begin{equation}
M^i = \left[c(\rb^i) + \delta c^i \eta^i\right]/\bar{c} + \Delta^i \label{eq:M} 
\end{equation}
where $\bar{c}$ is the mean concentration over the cluster, $\bar{c} = N^{-1} \sum_i c(\rb^i)$, $\eta^i$ are uncorrelated Gaussian noises with zero mean and unit variance and $\Delta^i$ are uncorrelated Gaussian noises with zero mean and variance $\sd^2$.  Stochastic fluctuations in ligand-receptor binding are taken into account in the term $\delta c^i$, where $(\delta c^i / c(\rb^i))^2 = \frac{1}{n_r} \frac{(c^i+K_D)^2}{c^i K_D}$.  This is the error in concentration sensing from a single snapshot of $n_r$ receptors with simple ligand-receptor kinetics and dissociation constant $K_D$ (\cite{kaizu2014berg}, Appendix \ref{app:conc}). 
\eq \ref{eq:M} assumes that cell-cell variance additively corrupts the measurement of the concentration $c(\rb)$ {\it after} an adaptation to the overall level of signal across the cluster $\bar{c}$. This is natural if the primary cell-to-cell variation is {\it downstream} of adaptation, as found to be a reasonable model in \cite{wang2012diverse}.

To determine gradient sensing accuracy, we perform maximum likelihood estimation (MLE) of $\gv$ in \eq \ref{eq:M}, as in past approaches for single cell gradient sensing \cite{hu2011geometry}. We obtain the MLE estimator $\gh$ numerically ({\it Methods}, Appendix \ref{app:mle}), and thus the uncertainty $\sg^2 \equiv \davg{|\hat{\gv} - \gv|^2}$  (\fig \ref{fig:ccv_illus}b, symbols), where $\langle \cdots \rangle$ is an average over CCV and ligand-receptor binding. For fixed and roughly circular (isotropic) cluster geometry, if the concentration change across the cluster is small, $g R_\textrm{cluster} \ll 1$, $\sg^2$ can be approximated by assuming $\delta c^i$ is constant across the cluster, resulting in
\begin{equation}
\davg{|\hat{\gv} - \gv|^2} \approx \frac{2}{\chi} \left(\sd^2 + \frac{1}{n_r} \frac{(\bar{c}+K_D)^2}{\bar{c} K_D} \right) \label{eq:combined_error}
\end{equation}
Here, $\chi = \frac{1}{2}\sum_i |\dri|^2$ is a shape parameter, and $\dri = \vb{r}^i - \vb{r}_{\textrm{cm}}$ is cell position relative to cluster center of mass.
Evaluating this expression reveals that it is an excellent approximation to the numerically-obtained uncertainty (dashed lines, \fig \ref{fig:ccv_illus}b).

The approximate expression for the uncertainty, Eq. \ref{eq:combined_error},
allows us to quantify the relative contribution of receptor-ligand fluctuations and CCV to the gradient sensing error. For background concentrations $\bar{c}$ near the receptor-ligand equilibrium constant $K_D$ and for typical receptor numbers in eukaryotic cells ($n_r \sim 10^5$ \cite{hesselgesser1998identification,macdonald2008heterogeneity}), $\delta c / \bar{c}$ can be smaller than 0.01. 
Protein concentrations, on the other hand, often vary between cells to 10\%-60\% of their mean \cite{niepel2009non} -- hence we estimate $\sd \approx 0.1-0.6$. Thus, we expect CCV to dominate gradient sensing error and that
 the error from concentration sensing and receptor binding can be neglected completely if $\sd > 0.1$ (\fig \ref{fig:ccv_illus}b). 
 \eq \ref{eq:combined_error} also reveals that 
 CCV masks the impact of changing background concentration. When $\sd = 0$, gradient sensing is limited by ligand-receptor fluctuations, and increases as $\bar{c}$ moves away from $K_D$ (\fig \ref{fig:ccv_illus}d) -- accuracy decreases if either few receptors are bound, or if receptors are saturated. As CCV increases, $\sg^2$ no longer depends strongly on $\bar{c}$ (\fig \ref{fig:ccv_illus}d). 
 Finally, Eq. \ref{eq:combined_error} shows that gradient sensing accuracy 
 depends on  the shape parameter $\chi$ and, therefore, on 
 cluster size. 
 For hexagonally packed clusters of cells with unit spacing\footnote{We measure in units of the cell diameter; see {\it Methods}}, a cluster with $Q$ layers has $N = 1 + 3Q + 3Q^2$ cells and $\chi(Q) = (5/8) Q^4 + (5/4) Q^3 + (7/8) Q^2 + (1/4) Q$ (Appendix \ref{app:Q}), i.e. $\chi(Q) \sim Q^4 \sim N^2$. 
Clusters of increasing size then have an error that decreases as $1/N^2$ (\fig \ref{fig:ccv_illus}c); {this scaling is similar to earlier results for single cells (Appendix \ref{app:Q}).}

\section{Reducing estimation error by time-averaging}
\label{sec:timeaverage}

\begin{figure}[ht!]
 \centering
\includegraphics[width=90mm]{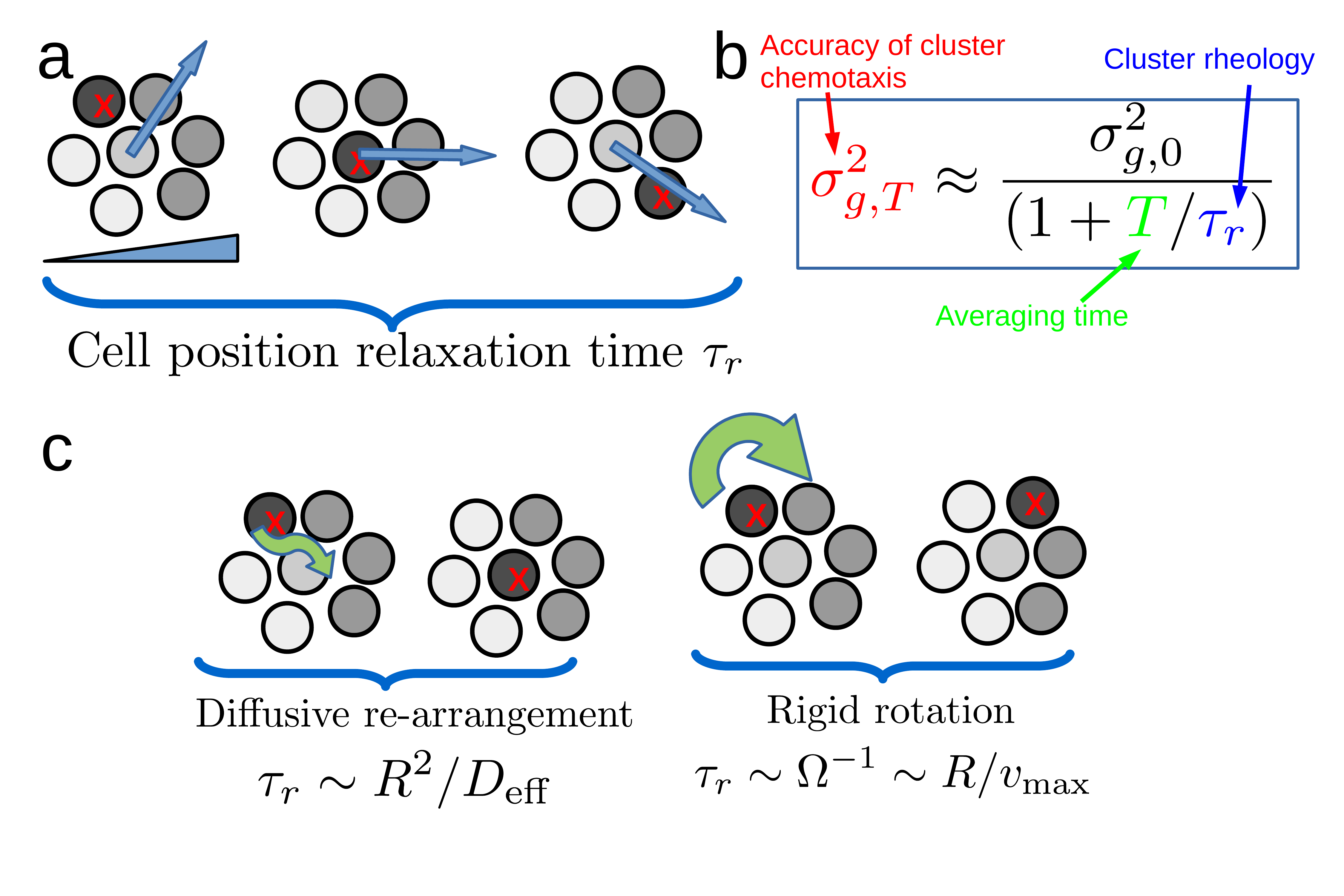}
 \caption{\linespread{1.0}\selectfont{}{\bf Time-averaging links fluidity and accuracy}. a) Schematic drawing of how cell-cell rearrangement can change bias due to CCV. Shades of gray indicate measured signal $M$; a cell with strong response (marked with $X$) moves through the cluster, leading to biases in gradient estimate (blue arrow). The characteristic relaxation time for this bias is $\tau_r$ (see text). b) This leads to a link between the timescale $\tau_r$, which is a measure of the cluster's rheology, and chemotactic accuracy $\sgt$ (box). c) Different re-arrangement mechanisms will depend on cluster size in different ways (see text and Appendix \ref{app:mechanisms}). }
 \label{fig:timeaverage}
 \end{figure}

If a cluster made $n$ independent measurements, it could reduce $\sg^2$ by a factor of $n$. In single-cell gradient sensing, independent measurements can be made by averaging over time -- improving errors by a factor $\sim T/\tau_{\textrm{corr}}$, where $T$ is the averaging time, and $\tau_{\textrm{corr}}$ the measurement correlation time. At first glimpse, time averaging seems unlikely to help with CCV, when correlation times for protein levels can be longer than cell division times, reaching 48 hours in human cells \cite{sigal2006variability}. However, since gradient sensing bias from CCV depends on the locations of strong- and weak-signaling cells within the cluster, time averaging can be successful if it is over a time long enough for the cluster to re-arrange. This is true even if, as we initially assume, CCV biases $\Delta^i$ are time-independent. We expect gradient sensing error with time averaging, $\sgt^2$, will decrease by a factor of $T/\tau_r$ from $\sgz^2$, where $\tau_r$ is a correlation time related to cell positions (\fig \ref{fig:timeaverage}). Is this true, and how should we define $\tau_r$?

Our earlier results suggest that CCV dominates the gradient sensing error. Ligand-receptor noise will also be even less relevant in the presence of time averaging, as the receptor relaxation time (seconds to minutes \cite{wang2007quantifying}) is much faster than that for cluster re-arrangement (tens of minutes or longer). We therefore completely neglect ligand-receptor binding fluctuations, allowing an analytical solution for the MLE estimator $\gh$ ({\it Methods}, Appendix \ref{app:mle}).

How much does time-averaging reduce error? If we average the MLE estimate $\hat{\gv}$ over a time $T$ by applying a kernel $K_T(t)$, i.e., we define $\ght(t) \equiv \int_{-\infty}^{\infty} \gh(t') K_T(t-t') dt'$ and $\sgt^2 \equiv \davg{|\ght-\vb{g}|^2}$, we can derive (Appendix \ref{app:timeaverage})
\begin{equation}
\sgt^2 = \sgz^2 \times \int_{-\infty}^\infty \frac{d\omega}{2\pi} |K_T(\omega)|^2 C_{rr}(\omega) \label{eq:snrt}
\end{equation}
where $C_{rr}(t'-t'') \equiv \langle \boldsymbol\delta\rb(t')\cdot\boldsymbol\delta\rb(t'') \rangle / \langle |\boldsymbol\delta\rb|^2 \rangle$ is the normalized cell position-position correlation function, $C_{rr}(\omega)$ its Fourier transform, and $\sgz^2 = 2 \sd^2/\chi$ is the error in the absence of time-averaging. To derive \eq \ref{eq:snrt}, we make two approximations: 1) the cluster has a constant and isotropic shape, and 2) re-arrangement of cell positions relative to the center of mass is independent of the particular values of $\Delta$.
 The first approximation is not necessary, but is a useful simplification; a generalized result is given in Appendix \ref{app:timeaverage}. {The second approximation  assumes that averaging over CCV and averaging over cell positions are independent. This decoupling approximation is necessary to characterize cluster fluidity and mechanics separately from the details of signaling. It excludes, e.g. models where cells with larger-than-average $\Delta$ sort out from the cluster. We will discuss potential errors due to this approximation later in the paper.}

For exponential position-position correlation functions and averaging, $C_{rr}(t) = \exp(-t/\tau_r)$ and $K_T(t) = \theta(t) \frac{1}{T} e^{-t/T}$, where $\theta(t)$ is the Heaviside step function, \eq \ref{eq:snrt} is simple:
\begin{equation}
\sgt^2 = \frac{\sgz^2}{1+T/\tau_r} \label{eq:snrt_simp}
\end{equation}
In other words, gradient sensing accuracy can be improved by taking 
$T/\tau_r$ independent measurements in a time $T$.
Crucial in this reduction is the position-position correlation time $\tau_r$ which depends on the cluster rearrangement mechanism. 
Two natural mechanisms are persistent cluster rotation and neighbor re-arrangements within the cluster (\fig \ref{fig:timeaverage}c). These mechanisms may coexist, as when cells slide past one another during cluster rotation \cite{rappel1999self}. 
$\tau_r$ can depend on cluster size; for diffusive rearrangements, we expect $\tau_r \sim R^2/D_{\textrm{eff}}$, and for persistently rotating clusters, $\tau_{\textrm{r}} \sim R/v_{\textrm{cell}}$ (Appendix \ref{app:mechanisms}). 

We have assumed that the CCV is time-independent over our scale of interest -- consistent with the long memory found in \cite{sigal2006variability}. If $\Delta$ changes faster than the cluster re-arranges, our results can be straightforwardly modified. Generalizations of \eq \ref{eq:snrt} and \eq \ref{eq:snrt_simp} to this case are provided in Appendix \ref{app:timeaverage}. 

\begin{figure*}
 \centering
\includegraphics[width=180mm]{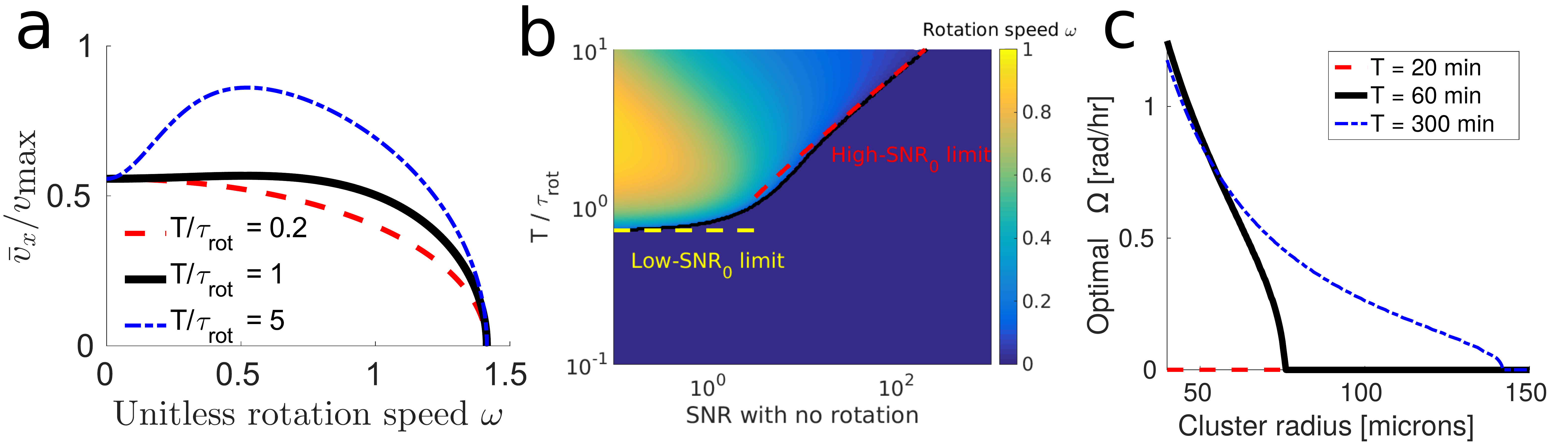}
 \caption{\linespread{1.0}\selectfont{}{\bf Cluster rotation can improve directed cluster motility}.  {\bf a}: As the averaging time $T$ is increased above the characteristic rotational timescale $\tau_\textrm{rot} = R/v_{\textrm{max}}$, the mean cluster velocity in the gradient direction $\bar{v}_x$ is maximized for nonzero rotational speed. $\snr_0 = 1$ in this panel. $\omega = \Omega \tau_\textrm{rot}$ is the unitless rotational speed. {\bf b}: Rotation improves chemotaxis at long averaging times $T$ and low $\snr_0$ (bad gradient sensing in the absence of rotation). Color map shows the value of $\omega$ that maximizes $\davg{v_x}$, found by numerical evaluation; black line shows the $\omega \approx 0$ contour. {\bf c}: Cluster rotation is preferred at small cluster radii. In this graph, $\snr_0$ is estimated by using $\chi \approx \frac{\pi}{4} \rho_c R^4$, where $\rho_c \approx 3.2 \times 10^{-3} \um^{-2}$ is the number of cells per unit area in the cluster ({\it Methods}).}  
 \label{fig:rot}
 \end{figure*}

\section{Tradeoffs in collective accuracy and motility: cluster rotation}
\label{sec:rotate}

Our central result (\eq \ref{eq:snrt}) shows that clusters can improve their chemotactic accuracy by changing cell positions. The simplest mechanism to do this is cluster rotation, which occurs in border cell clusters \cite{combedazou2016myosin} and transiently in leukocyte clusters \cite{malet2015collective}. When should a cluster  actively rotate in order to increase its accuracy? Rotation creates an important tradeoff: more work must be put into rotating, and therefore less into crawling up the gradient \footnote{We note that this is most relevant if motility is a large portion of the cluster's energy budget, a complex, cell-type-dependent question \cite{purcell1977life,flamholz2014quantified,katsu2009substantial}}

For constant work of motility, the maximum speed of a cluster of radius $R$ that rotates with angular speed $\Omega$ is ({\it Methods})
\begin{equation}
v(\Omega) = v_\textrm{max}\sqrt{1 - \frac{1}{2}\Omega^2\tau_{\textrm{rot}}^2} \label{eq:rot_tradeoff_main}
\end{equation}
where $\tau_\textrm{rot} = R/v_\textrm{max}$ and $v_\textrm{max}$ the maximum cluster speed absent rotation. The cluster cannot rotate faster than $\Omega_\textrm{max} = \sqrt{2}/\tau_\textrm{rot}$. 

If a cluster follows its best estimate $\ght$ with speed $v(\Omega)$ given by \eq \ref{eq:rot_tradeoff_main}, it can improve its velocity in the gradient direction by rotating when the averaging time $T$ is long compared with $\tau_\textrm{rot}$ (\fig \ref{fig:rot}a) \footnote{The cluster's directionality is always improved by rotating, so there is no tradeoff unless speed of motion matters.} We find that the optimal rotation speed $\Omega$ that maximizes the upgradient speed depends only on the signal-to-noise ratio without rotation, $\snr_0 \equiv \frac{1}{2}g^2/\sigma_{\gv,0}^2$ and $T/\tau_\textrm{rot}$ (\fig \ref{fig:rot}b, {\it Methods}). Mammalian cells have speeds in the range of microns/minute and radii of tens of microns, so to benefit from averaging ($T \gtrsim\tau_\textrm{rot}$), $T$ must be longer than tens of minutes. The timescale $\tau_\textrm{rot}$ and the signal-to-noise-ratio $\snr_0$ both depend on cluster size -- larger clusters with more cells are both better gradient sensors and more difficult to drive to large angular speeds. As a consequence, the optimal $\Omega$ is highly cluster-size-dependent: as cluster size decreases, there is a continuous transition to nonzero optimal $\Omega$ if $T$ is sufficiently long (\fig \ref{fig:rot}c).

\begin{figure*}[htb]
 \centering
\includegraphics[width=170mm]{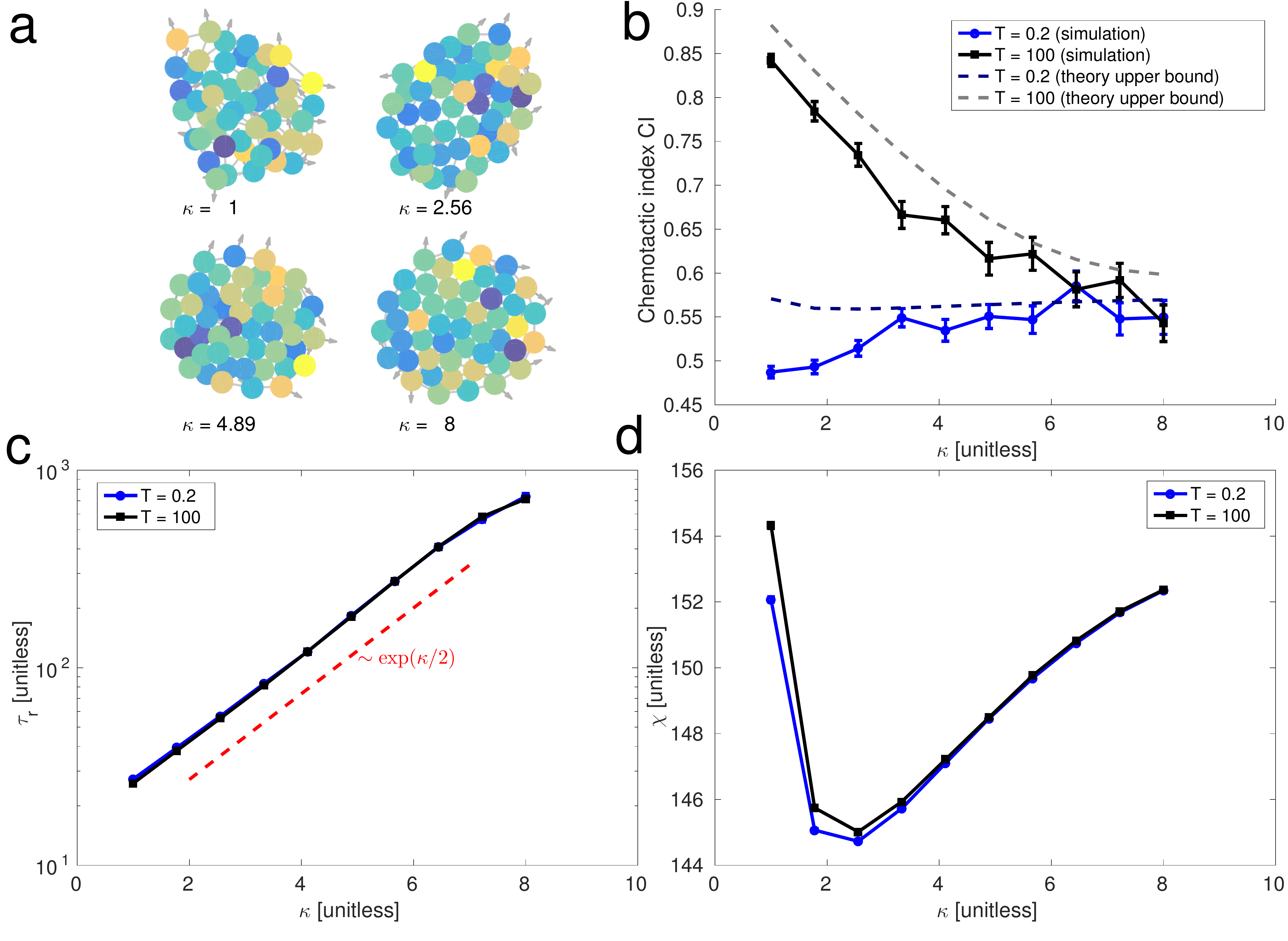}
 \caption{\linespread{1.0}\selectfont{}{\bf Cluster fluidity increases cluster accuracy}. {\bf a:} Typical configurations of cell clusters (plotted from simulations with $T = 0.2$). Color indicates measured signal $M^i$, lines connect neighboring cells, and the arrows indicate the polarity $\vb{p}$. {\bf b:} When time-averaging is significant, chemotactic index of clusters decreases as cell-cell adhesion stiffness is increased; this behavior is captured by the upper bound computed using \eq \ref{eq:snrt}. {\bf c:} Positional relaxation time $\tau_r$ decreases with stiffness $\kappa$ roughly as $\tau_r \sim e^{\kappa/2}$; $\tau_r$ does not strongly depend on averaging time $T$. {\bf d:} Cluster size parameter $\chi$ is not strongly dependent on $\kappa$ in this range of $\kappa$. All plots are computed by averaging over 600 simulations of $N = 50$ cells, each composed of $2\times10^4$ timesteps with $\Delta t = 0.02$. $\sd = 0.3$, $D_\psi = 1$, $\tau = 1$, $\ell = 1$, and $g = 0.025$. The first $2\times \textrm{max}(\tau,T)$ time units of the simulation are discarded, to allow the system to reach a steady-state.}
 \label{fig:kappa}
 \end{figure*}

\section{Linking chemotactic accuracy and fluidity}
\label{sec:fluidity}

Clusters with more cell rearrangement are more accurate by \eq \ref{eq:snrt_simp}. To further quantify the consequences of cell rearrrangements, we model a cluster of cells as self-propelled particles that follow the cluster estimate $\ght$ with a noise characterized by angular diffusion $D_\psi$ and with cell-cell connections modeled as springs of strength $\kappa$ between Delaunay neighbors ({\it Methods}). We emphasize that the angular diffusion parameterized by $D_\psi$ is an additional source of noise: as $D_\psi$ increases, cells are less accurate in following the cluster's estimate of the gradient. 
These two parameters are systematically varied to study the 
effects of cluster fluidity on chemotactic accuracy. 

\subsection{Cluster fluidity improves cluster chemotaxis}
Increasing cell-cell adhesion $\kappa$ makes clusters more ordered, moving between fluidlike and crystalline states (\fig \ref{fig:kappa}a). As a consequence, rearrangement slows significantly (\fig \ref{fig:kappa}c) with $\tau_r \sim \exp(\kappa/2)$ ($C_{rr}(t)$ is single-exponential). 

Cluster structure and size change when clusters fluidize (\fig \ref{fig:kappa}a), which may in principle affect the shape parameter $\chi$, which also strongly affects the chemotactic accuracy (\eq \ref{eq:combined_error}). However, in our simulations $\chi$ is not significantly dependent on $\kappa$, changing by under 10\% (\fig \ref{fig:kappa}d). Averaging time $T$ also has only a weak effect on cluster shape and dynamics -- changes in $\tau_r$ and $\chi$ when the averaging time $T$ is increased by orders of magnitude are small (\fig \ref{fig:kappa}). This is consistent with our assumption decoupling the gradient estimate and cell rearrangements, suggesting clusters should obey the bound \eq \ref{eq:snrt}. 

We can, using the results of Section \ref{sec:timeaverage}, predict the cluster chemotactic index, $\textrm{CI} \equiv \langle V_x / |\vb{V}| \rangle$, where $\vb{V}$ is the cluster velocity. Assuming $\vb{V} \sim \ght$, we can compute $\textrm{CI}$ from $\sgt$ given by \eq \ref{eq:snrt_simp} ({\it Methods}). This requires parameters $\tau_r$ and $\chi$ (measured from simulations), and $\vb{g}$, $T$, and $\sd$ (known).  {We note that our approach, which extracts $\tau_r$ and $\chi$ from cell trajectories, could also be applied to experimental data; in that case, $\vb{g}$ would still be known, but the extent of time-averaging ($T$) and the error due to CCV ($\sd$) would have to be determined by fitting to the data.} This prediction should be an upper bound to the measured $\textrm{CI}$, because our model includes additional noise beyond the assumptions of \eq \ref{eq:snrt_simp}, via $D_\psi$. 
As expected, cluster CI decreases significantly as clusters solidify and the relaxation time $\tau_r$ increases. The simulation data qualitatively follows the predicted
upper bound (\fig \ref{fig:kappa}b).
When the averaging time $T$ is reduced below typical relaxation times, the CI significantly decreases. In addition, for this short time-averaging, changing cluster stiffness no longer strongly affects CI. 

\subsection{Increasing single-cell stochasticity can increase cluster accuracy}

\begin{figure*}[htb]
 \centering
 \includegraphics[width=180mm]{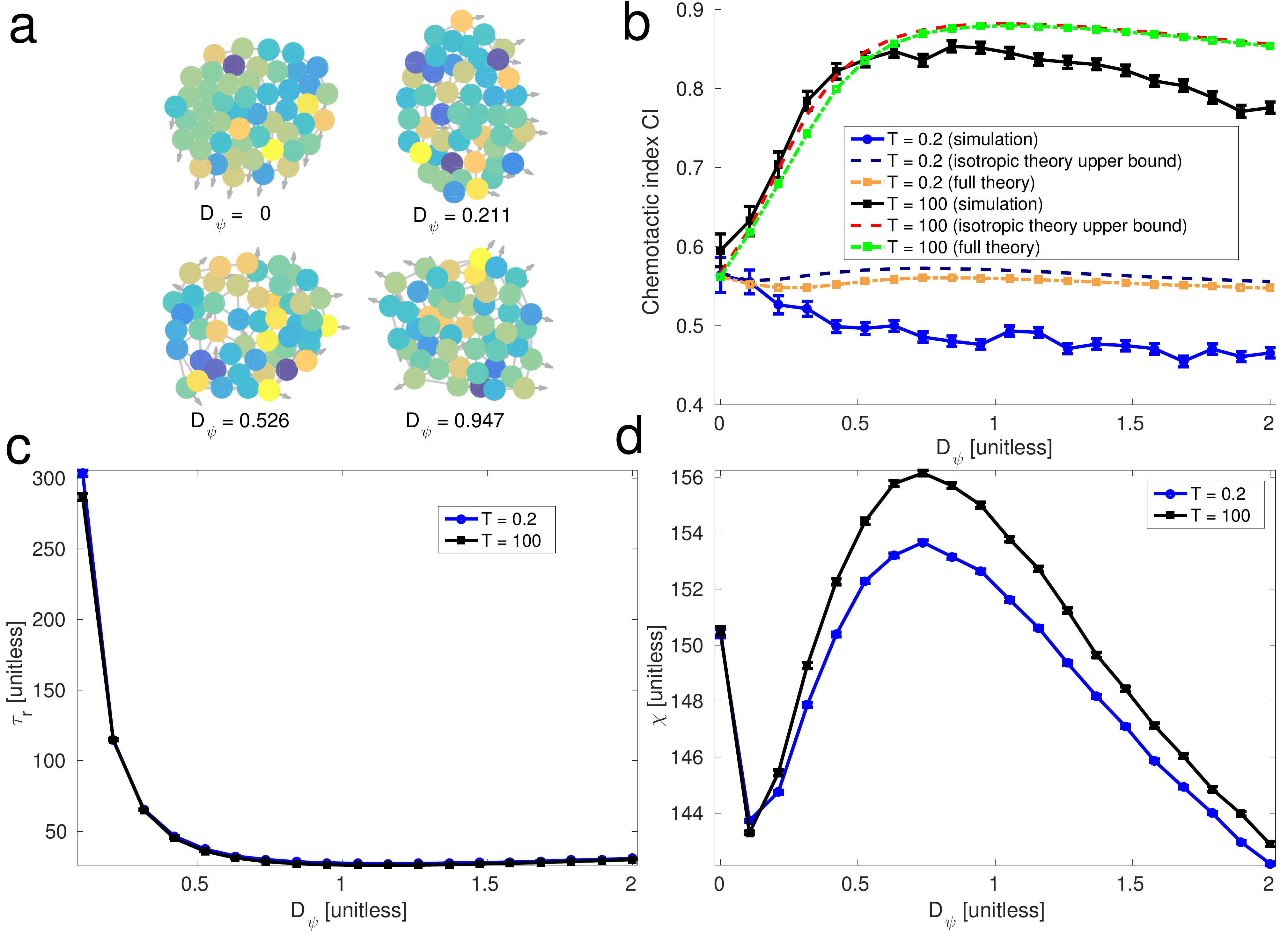}
 \caption{\linespread{1.0}\selectfont{}{\bf Finite levels of single-cell stochasticity increase accuracy}. {\bf a:} Typical configurations of cell clusters (plotted from simulations with $T = 100$). Color indicates measured signal $M^i$, lines connect neighboring cells, and the arrows indicate the polarity $\vb{p}$. {\bf b:} When time-averaging is significant, chemotactic index of clusters first increases as increasing single-cell noise $D_\psi$ fluidizes the cluster, then falls below the upper bound computed using \eq \ref{eq:snrt}. We also plot an extended theory not assuming cluster isotropy, derived in Appendix \ref{app:timeaverage}. Simulations for small $D_\psi$ may slightly exceed the upper bound (see text). {\bf c:} Positional relaxation time $\tau_r$ first decreases, then increases as $D_\psi$ is increased; the point $D_\psi = 0$, which has $\tau_r$ effectively infinite, is not shown. $\tau_r$ does not strongly depend on averaging time $T$. {\bf d:} Cluster size parameter $\chi$ weakly depends on fluidization by $D_\psi$. All plots are computed by averaging over 600 simulations of $N = 50$ cells, each composed of $2\times10^4$ timesteps with $\Delta t = 0.02$. $\sd = 0.3$, $\kappa = 1$, $\tau = 1$, $\ell = 1$, and $g = 0.025$. The first $2\times \textrm{max}(\tau,T)$ time units of the simulation are discarded, to allow the system to reach a steady-state.}
 \label{fig:Dpsi}
 \end{figure*}

Any mechanism that fluidizes the cluster can decrease the correlation time $\tau_r$. Because of this, {\it increasing} noise can improve cluster chemotactic accuracy (\fig \ref{fig:Dpsi}). We increase single-cell angular noise $D_\psi$, and see an initial sharp increase in cluster chemotactic index as $D_\psi > 0$ (\fig \ref{fig:Dpsi}b). At larger values of $D_\psi$, cluster CI decreases below the bound set by \eq \ref{eq:snrt_simp}, as the additional noise added degrades the gradient-following behavior. Without significant time-averaging ($T = 0.2$), additional noise primarily impedes chemotactic accuracy.  

Why can extra noise $D_\psi$ help sensing? For $D_\psi = 0$, all cells follow the best estimate $\ght$ precisely, leading to an ordered cluster (\fig \ref{fig:Dpsi}a) with $\tau_r$ effectively infinite. As $D_\psi$ is increased, the cluster fluidizes, and the relaxation time decreases strongly (\fig \ref{fig:Dpsi}c) resulting in more independent measurements. As in \fig \ref{fig:kappa}, this fluidization is only relevant if the averaging time $T$ exceeds the relaxation time, so when $T = 0.2$, the effect of increasing $D_\psi$ is solely detrimental to chemotaxis.

{In deriving our bound, we made two key approximations: cluster isotropy and decoupling. These approximations are exact for the rigid cluster rotation in Sec. \ref{sec:rotate}, but only approximate for this self-propelled particle model. As a consequence, at small $D_\psi$, simulated clusters have chemotactic indices slightly exceeding our predictions (\fig \ref{fig:Dpsi}b). This error likely arises from emergent couplings between cluster shape and $\Delta_i$ -- clusters may spread perpendicular to $\ght$, weakening the decoupling approximation (\fig \ref{fig:Dpsi}a). The approximation of cluster isotropy can be removed (Appendix \ref{app:timeaverage}), and does not resolve the violation of the bound (\fig \ref{fig:Dpsi}b). Despite this potential error source, the model captures CI variation over a broad range of parameters (Appendix \ref{app:params}).}

\section{Discussion}

Our study results in several predictions and suggestions for 
experiments that investigate collective chemotaxis. For example, we
predict that, when CCV is large, gradient sensing error is insensitive to background concentration (\fig \ref{fig:ccv_illus}c).
This is consistent with recent measurements 
on developing organoids that show that the up-gradient bias is not strongly dependent on mean concentration \cite{ellison2016cell}, though in contrast with results on lymphocyte clusters \cite{malet2015collective}.  
Furthermore,  
if CCV limits collective chemotaxis, clusters gradient sensing {\it in vivo} should have tightly regulated expression of proteins relevant to the signal response. Interestingly, measurements of zebrafish posterior lateral line primordium \cite{venkiteswaran2013generation} show tightly-controlled Sdf1 signaling, as measured by Cxcr4b internalization, suggesting that 
CCV may be small enough to allow for accurate gradient sensing.

We also show that there is a direct link between fluidity and chemotaxis 
as shown by \eq \ref{eq:snrt}.
Verifying this expression in experiments  requires simultaneous measurement of
several quantities, including cluster size, cluster re-arrangement, and signal gradient. Therefore, care has to be taken when modifying 
experimental conditions as these might change several of these 
quantities simultaneously. Altering adhesion, for example, changes both cluster fluidity and spreading as shown in a recent study 
using neural crest clusters \cite{kuriyama2014vivo}, creating a confounding factor.
Nevertheless, these types of experiments may be successful in setting bounds on possible time-averaging and the link between fluidity and chemotaxis.

Our results suggest that many recent experiments may need reinterpretation. Measured chemotactic accuracies can depend on cluster size \cite{theveneau2010collective,malet2015collective,cai2016modeling}; these results have been modeled without time averaging or CCV \cite{camley2016emergent,malet2015collective,cai2016modeling,camley2016collective}. Our results show that rearrangement times $\tau_r$ also influence chemotaxis -- and that $\tau_r$ depends on cluster size. Cluster relaxation dynamics are therefore an unexplored potential issue for interpreting collective gradient sensing experiments.

Essential in the reduction of gradient sensing errors due to CCV is the existence of 
a biochemical or mechanical memory that can perform a time average over tens of minutes. There are several possibilities. First, memory could be external to the cluster -- e.g. stored in extracellular matrix structure, or a long-lived trail \cite{lim2015neutrophil}. Secondly, supracellular structures like actin cables influence cell protrusion and leader cell formation \cite{reffay2014interplay}, suggesting that collective directional memory could be kept by regulating actin cable formation and maintenance. Third, memory may be kept at the individual cell level by cells attempting to estimate their own bias level $\Delta^i$ and compensating for it. This contrasts with our straightforward average of the collective estimate $\gh$, but could be an important alternative mechanism.

Our results are critical for understanding the ubiquitous phenomenon of collective gradient sensing. The importance of CCV provides a new design principle: CCV must either be tightly controlled or mitigated by time-averaging. We also established a surprising link between a central {\it mechanical} property of a cluster -- its rheology -- and its sensing ability. This connects mechanical transitions like unjamming \cite{bi2016motility} to sensing, opening up new areas of study. In addition, our results show cluster accuracy depends strongly on cluster rearrangement mechanism. Finally, our results show that noise in cell motility can be beneficial for collective sensing.

\section{Methods}
\subsection*{Maximum likelihood estimation of gradient direction in the presence of cell-cell variation and ligand-receptor noise}
We compute the maximum likelihood estimate of gradient direction given the measured signal at cell $i$, $M^i$, given by \eq \ref{eq:M}. If the cluster of cells is in a shallow linear gradient, with concentration $c_0$ at the cluster's center of mass $\rb_{\textrm{cm}} = N^{-1} \sum_i \rb^i$, then $c(\rb) = c_0 \left[1 + \vb{g} \cdot (\rb-\rb_{\textrm{cm}}) \right]$ and thus $\bar{c} = c_0$. $M_i$ is then
$M^i = 1 + \gv\cdot\dri + (\delta c^i/c_0) \eta^i + \Delta^i$
with $\dri = \rb-\rb_{\textrm{cm}}$ and $(\delta c^i/c^i)^2 = \frac{1}{n_r} \frac{(c^i+K_D)^2}{c^i K_D}$, i.e. $(\delta c^i/c_0)^2 = \frac{1}{n_r} (1+\gv\cdot\dri+K_D/c_0)^2\frac{1+\gv\cdot\dri}{K_D/c_0}$. 

$\Delta^i$ are uncorrelated between cells, with a Gaussian distribution of zero mean and standard deviation \sd, i.e. $\davg{\Delta^i\Delta^j} = \sd^2 \delta^{ij}$ with $\delta^{ij}$ the Kronecker delta function. As $\eta^i$ and $\Delta^i$ are both Gaussian, the sum of these variables is also Gaussian, and the likelihood of observing a configuration of measured signals $\{M^i\}$ as $\like(\gv;\Mconf) = P(\Mconf | \gv)$, where $P(\Mconf | \gv)$ is the probability density function of observing the configuration \Mconf given parameters $\gv$.
\begin{equation}
\like(\gv;\Mconf) = \prod_{i} \frac{1}{\sqrt{2\pi h^i}} \exp\left[-\frac{(M^i - \mu^i)^2}{2 h^i} \right] 
\end{equation}
where $\mu^i = 1 + \gv\cdot\dri$ is the mean value of $M^i$ and $h^i = (\delta c^i / c_0)^2 + \sd^2$ its variance.  We want to apply the method of maximum likelihood by finding the gradient parameters $\hat{\gv}$ that maximize this likelihood, i.e.
\begin{equation}
\hat{\gv} = \textrm{arg} \, \max\limits_{\gv} \like(\gv;\Mconf)
\end{equation}
However, because of the complex dependence of $h^i$ on the gradient $\gv$, this is not possible analytically. We perform this optimization numerically using a Nelder-Mead method (Matlab's fminsearch), with an initial guess set by the maximum for $n_r \to \infty$ (i.e. neglecting concentration sensing noise), which can be found exactly. For numerical convenience, we maximize the log-likelihood $\ln \like(\gv;\Mconf)$, 
 $\ln \like(\gv;\Mconf) = -\frac{1}{2}\sum_{i} \ln h^i - \sum_i \frac{(M^i - \mu^i)^2}{2 h^i}$ 
up to an additive constant. 

In the limit of $n_r \to \infty$ (neglecting concentration noise), our model becomes a simple linear regression, and the log likelihood can be maximized analytically by finding \gh such that $\partial_{\gv} \ln \like(\gv; \Mconf)|_{\gh} = 0$ (Appendix \ref{app:mle}). The result is $\hat{\gv} = \mathcal{A}^{-1}\cdot \sum_i M^i \dri$, where $\mathcal{A}_{\alpha\beta} \equiv \sum_i\delta r^i_\alpha \delta r^i_\beta$. This estimator is simplest in the limit of roughly circular (isotropic) clusters, where $\sum_i (\delta x^i)^2 \approx \sum_i (\delta y^i)^2 \gg \sum_i \delta x^i \delta y^i$.  In this case, $\hat{\gv} = (\chi^{-1}) \sum_i M^i\dri$ where $\chi = \frac{1}{2}\sum_i |\dri|^2$. 

\subsection*{Cluster rotation dynamics}
How much speed does a cluster lose by rotating? One possibility is to assume the power expended in generating motility is constant.  Consider a circular cluster propelling itself over a surface, with the cells having velocity $\vb{v}(\rb)$; we expect that the frictional force per unit area between the cluster and substrate will be $\vb{f}_{\textrm{drag}} = -\xi \vb{v}$, where $\xi$ is a friction coefficient with the substrate. If all of the power available for motility is going into driving the cluster over the substrate, then we can write:
$P = -\int d^2 r \, \vb{v}\cdot\vb{f}_\textrm{drag} = \xi \int d^2 r |\vb{v}|^2$. 
If the cluster is traveling as a rigid, circular cluster with its maximum possible velocity, $\vb{v} = v_{\textrm{max}} \hat{\vb{x}}$ then
$P = \xi \pi R^2 v_\textrm{max}^2 \equiv \gamma_t v_\textrm{max}^2$ 
where $\gamma_t = \xi \pi R^2$ is the translational drag coefficient of the cluster.  If, instead, the cluster puts its entire power into rigid-body rotation with $\vb{v}(\rb) = \Omega_{\textrm{max}} r (-\sin \theta,\cos \theta)$ (in polar coordinates), then 
$P = \xi \Omega_\textrm{max}^2 \int d^2 r r^2 = \xi\frac{\pi}{2} R^4 \Omega_{\textrm{max}}^2 \equiv \gamma_r \Omega_\textrm{max}^2$
where $\gamma_r = \frac{\xi \pi}{2} R^4$ is the rotational drag coefficient of the cluster.
In general, the power dissipated if the cluster is moving rigidly with velocity $\vb{v}$ and angular speed $\Omega$ is
$P = \gamma_t v^2 + \gamma_r \Omega^2$
and hence, we find that the speed $v(\Omega$) that a cluster rotating with angular velocity $\Omega$ is able to travel to obtain is
\begin{equation}
v(\Omega) = \sqrt{v_\textrm{max}^2 - \frac{\gamma_r}{\gamma_t} \Omega^2} \label{eq:tradeoff}
\end{equation}
This quantifies one reasonable tradeoff between speed and angular velocity for a cluster. If the power available for cell motility is a small amount of the cell's energy budget \cite{purcell1977life,flamholz2014quantified,katsu2009substantial},
other tradeoffs may be more important and additional modeling will be necessary. 

We consider a circular cluster traveling towards its best estimate of the gradient with speed $v(\Omega)$ given by \eq \ref{eq:tradeoff}, and traveling in the direction of the estimator $\ght$.  We can then determine when the cluster maximizes its mean velocity in the direction of the increasing gradient, which we choose to be $x$, $\davg{v_x}$, as a function of $\Omega$.  This average is
\begin{equation}
\davg{v_x} = \sqrt{v_\textrm{max}^2 - \frac{\gamma_r}{\gamma_t} \Omega^2} \times \davg{\frac{\ghtx}{|\ght|}}
\end{equation}
We know from our results above and in Appendix \ref{app:timeaverage} that, for a fixed configuration, $\ght$ has a Gaussian distribution with mean $\vb{g} = g \hat{\vb{x}}$ and variance given by \eq \ref{eq:snrt}. The average of $\frac{\ghtx}{|\ght|}$ depends only on the signal-to-noise ratio $\snr_T \equiv \frac{1}{2}(g^2/\sigma_{\vb{g},T}^2)$, with $\davg{\frac{\ghtx}{|\ght|}} = C(\snr_T^{-1/2})$. 

Given the angular velocity $\Omega$, we can work out the distribution of $\ght$ by \eq \ref{eq:snrt}.  We know $\rb(t)\cdot\rb(0) = |\rb(0)|^2\cos \Omega t$, and hence $C_{rr}(t) = \cos \Omega t$ and its Fourier transform $C_{rr}(\omega) = \pi \left[\delta(\omega-\Omega) + \delta(\omega+\Omega)\right]$, and thus $\sgt^2 = \sgz^2 \times |K_T(\Omega)|^2 = \sgz^2/(1+\Omega^2T^2)$. 

By rescaling to unitless parameters, we then find that 
\begin{equation}
\davg{v_x}/v_\textrm{max} = \sqrt{1 - \frac{1}{2} \omega^2} \times C(\left[\snr_0 (1+\omega^2\widetilde{T}^2)\right]^{-1/2})
\end{equation}
where $\omega = \Omega R / v_{\textrm{max}}$ is the unitless angular velocity, $\snr_0 = \sd^{-2} g^2 \chi$ is the usual SNR with no averaging, and $\widetilde{T} = T v_{\textrm{max}}/R$ is the ratio of the averaging time to the characteristic rotational time $R/v_{\textrm{max}}$, and $C(\sigma)$ is the function given by \eq \ref{eq:csigma}.  When $\snr_0$ is sufficiently small, and $\widetilde{T}$ sufficiently large, $\davg{v_x}/v_\textrm{max}$ has a maximum at finite $\omega$ (\fig \ref{fig:rot}).

In the limit of low \snr, $C(\sigma) \approx \sqrt{\pi/8} \sigma^{-1}$, and we find
$\davg{v_x}/v_\textrm{max}$ is maximized by $\omega = \pm\sqrt{1-\frac{1}{2}\widetilde{T}^{-2}}$ when $\widetilde{T} > 1/\sqrt{2}$ and $\omega = 0$ otherwise.  For the large $\snr$ limit, $C(\sigma) \approx 1-\sigma^2/2$ and 
rotation will increase the mean velocity in the direction of the gradient when $\widetilde{T}^2 > \snr_0/2-1/4$.  More generally, it is possible to find the value of $\omega$ that maximizes $\davg{v_x}$ numerically.  We show the complete phase diagram in \fig \ref{fig:rot}b.

\subsection*{Particle-based model of collective cell migration}
We use a minimal model of collective cell migration, describing cells as self-propelled particles connected by springs:
\begin{align}
\label{eq:velocity}
\frac{d}{dt}\vb{r}^i &= \vb{p}^i + \sum_{j\sim i} \vb{F}^{ij} \\
\vb{p}^i &= (\cos\theta^i,\sin\theta^i) \\
\theta^i &= \arctan(\hat{g}_{T,y}/\hat{g}_{T,x}) +\psi^i \\
\frac{d}{dt} \psi^i &= -\tau^{-1} \sin \psi^i+ \sqrt{2 D_\psi} \xi^i(t) \label{eq:psi}
\end{align}
where $\xi^i(t)$ is a Gaussian Langevin noise with zero mean and $\langle \xi^i(t)\xi^j(t') \rangle = \delta^{ij}\delta(t-t')$, with $\delta^{ij}$ the Kronecker delta. In this model, the orientation of an individual cell $\theta^i$ is the cluster's best estimate of the gradient direction, $\arctan(\hat{g}_{T,y}/\hat{g}_{T,x})$, plus a noise $\psi^i$ which varies from cell to cell. $\tau$ here controls the persistence of this noise, and $D_\psi$ its amplitude; when $D_\psi$ is increased, each individual cell is worse at following the estimate $\ght$. The cell-cell forces are
\begin{equation}
\vb{F}^{ij} = -\kappa (|\dij-\ell|)\rij
\end{equation}
where $\dij = |\vb{r}^i-\vb{r}^j|$ and $\rij = (\vb{r}^i-\vb{r}^j)/\dij$.  The forces are only between neighboring cells $j\sim i$, where we define neighboring cells as any cells connected by the Delaunay triangulation of the cell centers (\fig \ref{fig:kappa}a); this approach resembles that of \cite{meineke2001cell}. We use the Euler-Maruyama method to integrate Eqns. \ref{eq:velocity}-\ref{eq:psi}. 

\subsection*{Simulation units} We have chosen our parameters in the simulation and throughout the paper to be measured in units where the equilibrium cell-cell separation $\ell = 1$ (i.e. the cell diameter is unity), and the velocity of a single cell in the absence of cell-cell forces $\vb{v} = \vb{p} = (\cos \theta,\sin\theta)$ has unit magnitude. For, e.g. neural crest cells, the cell diameters are of order 20 microns, and the cell speeds on the order of microns/minute -- so a unitless time of $T$ corresponds to roughly $\textrm{20 minutes} \times T$ in real time. However, cell size and speed varies strongly from cell type to cell type, so we prefer to present these results in their unitless form so that they can be more easily converted.

\subsection*{Computing chemotactic indices}

\begin{figure}[htb]
 \centering
 \includegraphics[width=90mm]{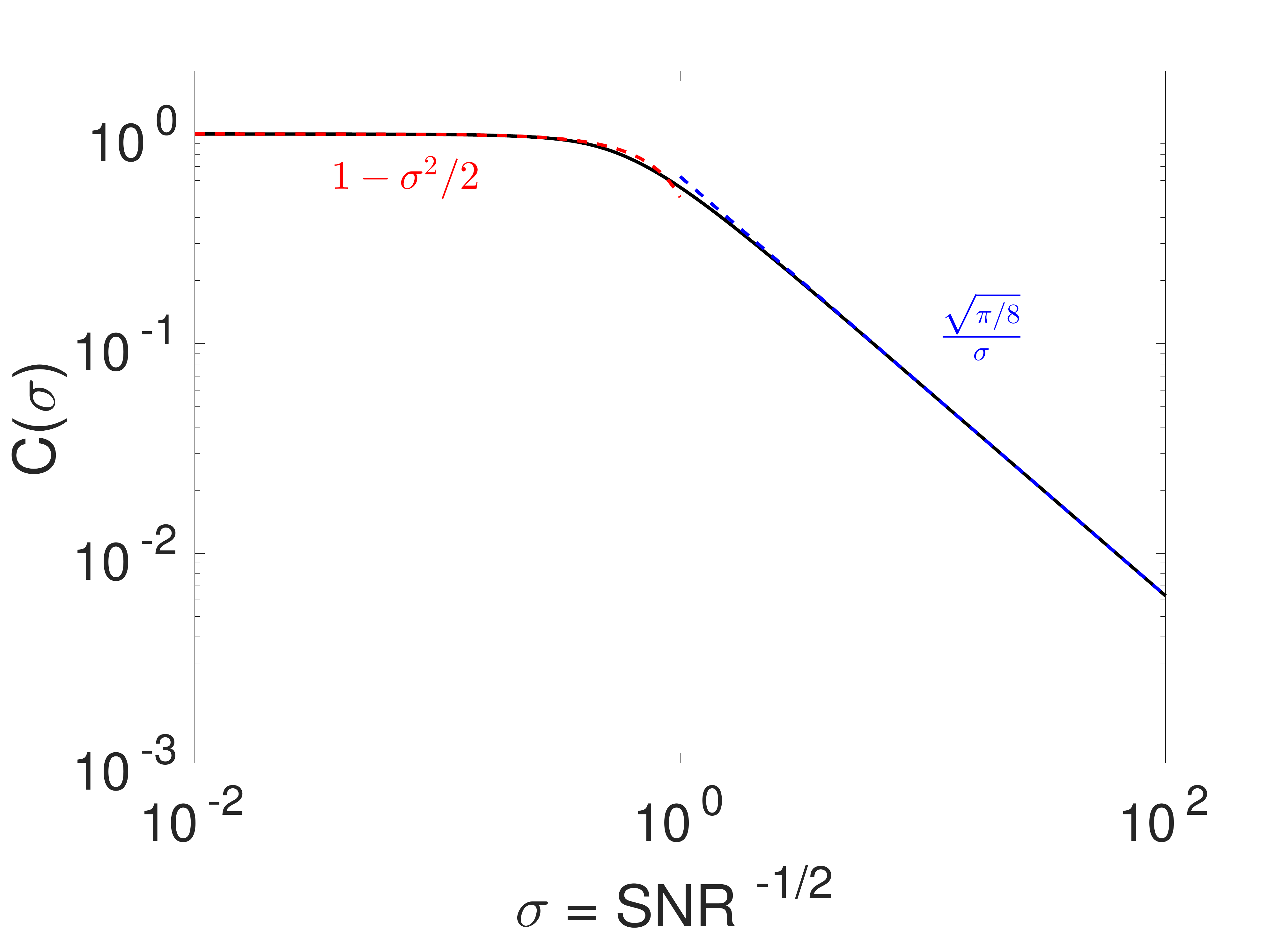}
 \caption{\linespread{1.0}\selectfont{}{\bf Relationship between instantaneous chemotactic index and $\snr$}. $C(\sigma)$ plotted numerically from definition in \eq \ref{eq:csigma}}
 \label{fig:csigma}
 \end{figure}

If we use the maximum likelihood method to make an estimate for the direction in which the cell moves, how do we translate between the uncertainty $\sg^2$ and the distribution of velocities?  We found that the MLE estimate for the gradient is $\gh = \vb{g} + \vb{\Lambda}$, with $\vb{\Lambda}$ a Gaussian random variable with zero mean and variance $\davg{\Lambda_x^2} = \davg{\Lambda_y^2} = \sg^2/2$ -- and similar results for the time-average $\ght$.  One measure of this estimate's accuracy is the {\it instantaneous chemotactic index} - or the cosine of the angle between the estimate and the gradient direction.  To compute this, if $\vb{g} = g \hat{\vb{x}}$ without loss of generality, we find
\begin{align}
\davg{\frac{\gh_x}{|\gh|}} &= \davg{\frac{g + \Lambda_x}{(g+\Lambda_x)^2 + \Lambda_y^2}} \\
&= \int \frac{dx dy}{2\pi\sigma} \frac{1 + x}{\left[(1+x)^2+y^2\right]^{1/2}} e^{-\frac{(x^2+y^2)}{2\sigma^2}} \\
&\equiv C(\sigma) \label{eq:csigma}
\end{align}
where $\sigma = \snr^{-1/2}$, with $\snr = \frac{1}{2} (g^2/\sg^2)$. These results carry over naturally to the time-averaged case if $\ght$ remains Gaussian -- we find $\davg{\frac{\gh_{T,x}}{|\ght|}} = C(\snr_T^{-1/2})$.  

The integral for $C(\sigma)$ can't be solved analytically, but we can find asymptotic forms for $C(\sigma)$ or evaluate it numerically. For $\sigma \gg 1$, we find $C(\sigma) \approx \sqrt{\pi/8} \sigma^{-1}$, and $C(\sigma) \approx 1 - \frac{1}{2} \sigma^2$ for $\sigma \ll 1$.

\section*{Acknowledgments}
We would like to thank Albert Bae and Monica Skoge for useful discussions, and many scientists from the Gordon Research Conference on Directed Cell Motility for interesting questions and reference suggestions. BAC also thanks Kristen Flowers for several useful suggestions.
This work was supported by NIH Grant No. P01 GM078586.

%

\clearpage
\setcounter{equation}{0}
\renewcommand*{\thefigure}{S\arabic{figure}}
\renewcommand*{\thetable}{S\arabic{table}}
\renewcommand*{\theequation}{S\arabic{equation}}

\onecolumngrid
\appendix
\begin{center}
{\Huge SI Appendix}
\end{center}
\section{Review of concentration sensing accuracy}
\label{app:conc}
With simple ligand-receptor kinetics, i.e. an on-rate of $k_\textrm{on} = k_+ c$ and an off-rate of $k_\textrm{off} = k_i$, the mean probability that each receptor will be occupied is $P_\textrm{on} = k_\textrm{on}/(k_\textrm{on}+k_\textrm{off}) = c(\rb)/(c(\rb)+K_D)$, where $K_D = k_-/k_+$ is the dissociation constant.  The variance in the occupation probability for an individual receptor is then $P_\textrm{on} - P_\textrm{on}^2 = c(\rb) K_D / (c(\rb)+K_D)^2$. By the central limit theorem, as the number of receptors $n_r$ becomes large, the number of {\it occupied} receptors on a cell will be a Gaussian distribution with mean $\bar{n} = n_r c(\rb)/(c(\rb)+K_D)$ and variance $\delta n^2 = n_r c(\rb) K_D / (c(\rb)+K_D)^2$. Translating this number of occupied receptors into an uncertainty in the local concentration via $\delta c = \frac{dc}{d\bar{n}} \delta n = \frac{(c+K_D)^2}{K_D n_r} \delta n$ \cite{kaizu2014berg}, we find $(\delta c / c)^2 = \frac{1}{n_r} \frac{(c+K_D)^2}{c K_D}$. 

\section{Maximum likelihood estimates of gradient direction via collective guidance in the presence of cell-cell variation and ligand-receptor noise}
\label{app:mle}

We begin with our model for the measured signal at cell $i$, $M^i$:
\begin{equation}
M^i = \left[c(\rb^i) + \delta c^i \eta^i\right]/\bar{c} + \Delta^i 
\end{equation}
where $\bar{c}$ is the mean concentration over the cluster of cells, $\bar{c} = N^{-1} \sum_i c(\rb^i)$. If we assume that the cluster of cells is in a shallow linear gradient, with the concentration measured at the cluster's center of mass $\rb_{\textrm{cm}} = N^{-1} \sum_i \rb^i$ being $c_0$, we have $c(\rb) = c_0 \left[1 + \vb{g} \cdot (\rb-\rb_{\textrm{cm}}) \right]$ and thus $\bar{c} = c_0$. We can then write the measured signal $M^i$ as
\begin{equation}
M^i = 1 + \gv\cdot\dri + (\delta c^i/c_0) \eta^i + \Delta^i 
\end{equation}
with $\dri = \rb-\rb_{\textrm{cm}}$ and $(\delta c^i/c^i)^2 = \frac{1}{n_r} \frac{(c^i+K_D)^2}{c^i K_D}$, i.e. $(\delta c^i/c_0)^2 = \frac{1}{n_r} (1+\gv\cdot\dri+K_D/c_0)^2\frac{1+\gv\cdot\dri}{K_D/c_0}$. 

We have assumed that $\Delta^i$ are uncorrelated between cells, with a Gaussian distribution of zero mean and standard deviation \sd, i.e. $\davg{\Delta^i\Delta^j} = \sd^2 \delta^{ij}$ with $\delta^{ij}$ the Kronecker delta. $M^i$, as the sum of the Gaussian variables $\eta^i$ and $\Delta^i$, is also Gaussian, and we can then write the likelihood of observing a configuration of measured signals $\{M^i\}$ as $\like(\gv;\Mconf) = P(\Mconf | \gv)$, where $P(\Mconf | \gv)$ is the probability density function of observing the configuration \Mconf given parameters $\gv$.
\begin{equation}
\like(\gv;\Mconf) = \prod_{i} \frac{1}{\sqrt{2\pi h^i}} \exp\left[-\frac{(M^i - \mu^i)^2}{2 h^i} \right] 
\end{equation}
where $\mu^i = 1 + \gv\cdot\dri$ is the mean value of $M^i$ and $h^i = (\delta c^i / c_0)^2 + \sd^2$ its variance.  We want to apply the method of maximum likelihood by finding the gradient parameters $\hat{\gv}$ that maximize this likelihood, i.e.
\begin{equation}
\hat{\gv} = \textrm{arg} \, \max\limits_{\gv} \like(\gv;\Mconf)
\end{equation}
However, because of the complex dependence of $h^i$ on the gradient $\gv$, analytically maximizing the likelihood is intractable. We instead perform this optimization numerically using a Nelder-Mead method (Matlab's fminsearch), with an initial guess set by the maximum for $n_r \to \infty$ (i.e. neglecting concentration sensing noise), given by \eq \ref{eq:ghat_simple}. For numerical stability and convenience, we will usually instead maximize the log-likelihood $\ln \like(\gv;\Mconf)$, which is
\begin{equation}
  \ln \like(\gv;\Mconf) = -\frac{1}{2}\sum_{i} \ln h^i - \sum_i \frac{(M^i - \mu^i)^2}{2 h^i} 
\end{equation}
up to an additive constant. 

In the limit of $n_r \to \infty$ (neglecting concentration noise), our model becomes a simple linear regression, and the log likelihood can be maximized analytically by finding \gh such that $\partial_{\gv} \ln \like(\gv; \Mconf)|_{\gh} = 0$.  This straightforwardly yields
\begin{equation}
\sum_i (M^i - 1 - \gh\cdot \dri)\dri= 0
\end{equation}
or, if we write $\gh = \ghx \hat{\vb{x}} + \ghy \hat{\vb{y}}$,
\begin{equation}
\nonumber
\begin{pmatrix} \sum_{i} (\delta x^i)^2 & \sum_i \delta x^i \delta y^i \\ \sum_i \delta x^i \delta y^i & \sum_i (\delta y^i)^2 \end{pmatrix} \begin{pmatrix} \ghx \\ \ghy \end{pmatrix} = \begin{pmatrix} \sum_i (M^i - 1) \delta x^i \\ \sum_i (M^i - 1) \delta y^i \end{pmatrix}
\end{equation}

We define $\mathcal{A}$ to be a matrix with elements $\mathcal{A}_{\alpha\beta} = \sum_i \delta r_\alpha \delta r_\beta$, where $\alpha,\beta$ run over the Cartesian coordinates $x,y$,
\begin{equation}
\hat{\gv} = (\mathcal{A}^{-1})\cdot \sum_i (M^i - 1) \dri \label{eq:ghat_simple}
\end{equation}
where we note that, as $\sum_i \dri = 0$, we can also simply use $\hat{\gv} = (\mathcal{A}^{-1})\cdot \sum_i M^i\dri$. 

These estimators are simpler for roughly symmetric clusters, where $\sum_i (\delta x^i)^2 \approx \sum_i (\delta y^i)^2 \gg \sum_i \delta x^i \delta y^i$.  In this case, $\mathcal{A}_{\alpha\beta} \approx \chi \delta_{\alpha\beta}$, where $\chi = \frac{1}{2}\sum_i |\dri|^2$ and $\delta_{\alpha\beta}$ is the Kronecker delta, and 
\begin{equation}
\hat{\gv} = (\chi^{-1}) \sum_i (M^i - 1) \dri
\end{equation}

We can also compute the asymptotic covariance of these estimators, which arises from the Fisher information matrix
\begin{align}
\mathcal{I}_{\alpha\beta} &= \davg{\left(\frac{\partial \ln \like}{\partial g_{\alpha}}\right) \left(\frac{\partial \ln \like}{\partial g_{\beta}}\right)}\\
&= -\davg{\frac{\partial^2 \ln \like}{\partial g_{\alpha} \partial g_{\beta}}} \\
&= \sd^{-2} \sum_i \delta r^i_\alpha \delta r^i_\beta = \sd^{-2} \mathcal{A}_{\alpha\beta}
\end{align}
where we use $\davg{\cdots}$ to indicate the average over the cell-to-cell systematic errors $\Delta_i$.

The Fisher information matrix controls the variance of our maximum-likelihood estimator \cite{kay1993fundamentals}, which is given by
\begin{align}
\davg{(\gh - \vb{g})_\alpha (\gh - \vb{g})_\beta} &= \left(\mathcal{I}^{-1}\right)_{\alpha\beta} \\
&= \sd^2 (\mathcal{A}^{-1})_{\alpha\beta}
\end{align}
In particular, for $\sg^2 \equiv \davg{|\gh-\vb{g}|^2}$, 
\begin{equation}
\sg^2 = \sd^2 \textrm{tr} A^{-1} \label{eq:traceerr}
\end{equation}

In the symmetric cluster limit, the matrix is diagonal, and the result is simply $\sigma_{\alpha}^{2} = \sd^{2}/\chi$ where $\sigma_{\alpha}$ is the standard deviation of the estimator $\hat{g}_\alpha$, and hence $\sg^2 = \sigma_x^2 + \sigma_y^2$,
\begin{equation}
\sg^{2} = 2\sd^{2}/\chi \label{eq:sigx}
\end{equation}

\section{Computation of $\chi$ for cells in hexagonally-packed cluster}
\label{app:Q}

 \begin{figure}[ht!]
 \centering
 \includegraphics[width=90mm]{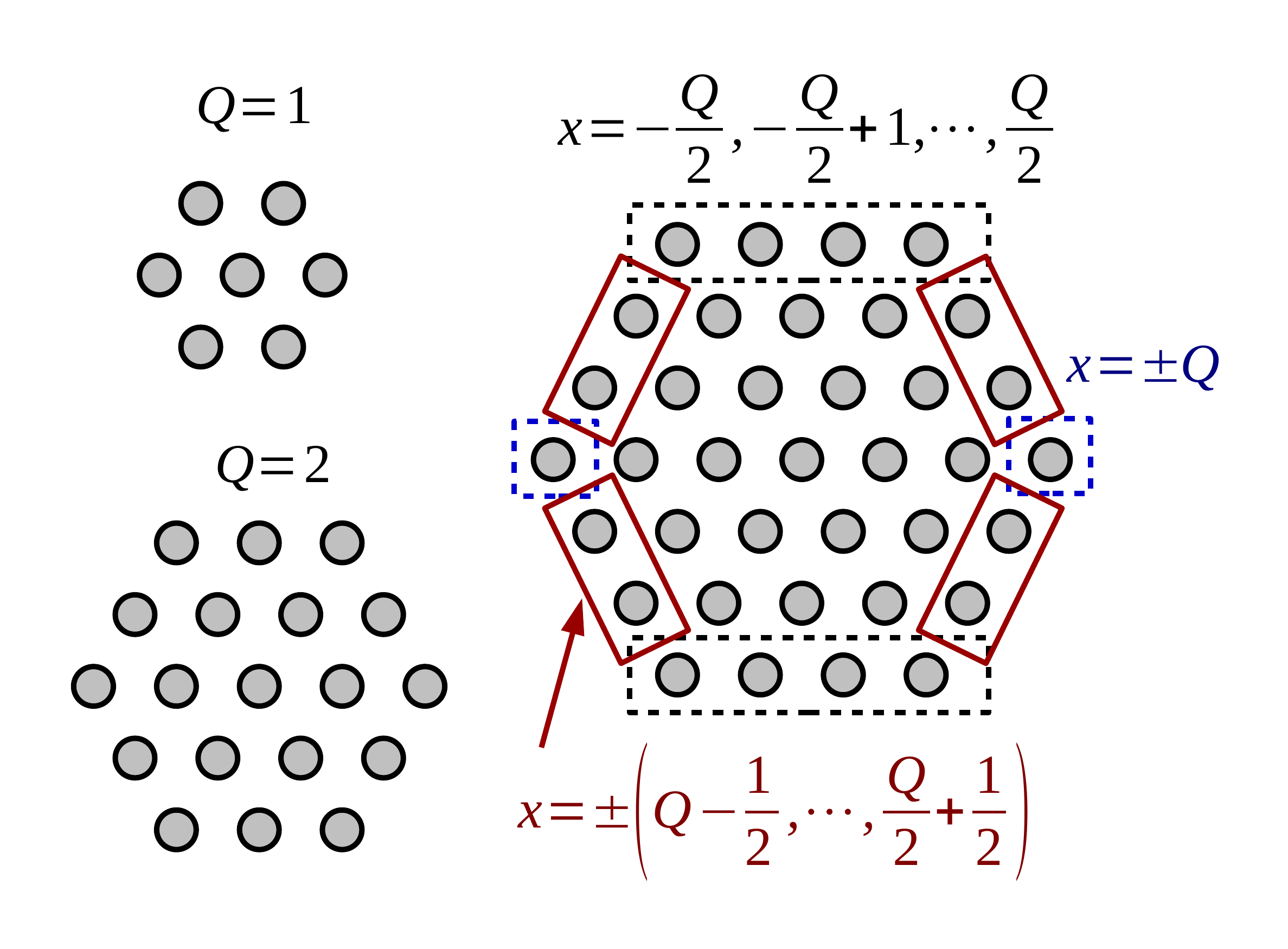}
 \caption{\linespread{1.0}\selectfont{} Illustration of $Q$-layer hexagonally packed cell clusters and computation of $\chi(Q)$.}  
 \label{fig:Q}
 \end{figure}

When we keep a fixed cluster geometry, we choose to work with hexagonally-packed cell clusters, following our earlier work \cite{camley2016emergent}. We illustrate clusters with $Q = 1, 2,$ and $3$ layers in \fig \ref{fig:Q}.

How can we calculate $\chi(Q)$? Because of the isotropy of the cluster, it is easiest to work with $\chi = \frac{1}{2} \sum |\dri|^2 = \sum (\delta x^i)^2 = \sum (x^i)^2$. We know that for a single cell ($Q = 0$), $\chi(0) = 0$. We can then determine $\chi(Q)$ in terms of $\chi(Q-1)$ by computing $x^2$ for each of the cells in the outer layer, which we call $G(Q) = \sum_{\textrm{i in outside layer}} (x^i)^2$. Then, $\chi(Q) = \chi(Q-1) + G(Q) = \sum_{q=1}^{Q} G(q)$. 

To compute $G(Q)$, there are three relevant portions of the cluster, as drawn in \fig \ref{fig:Q}. These are the two left and rightmost extreme cells at $x = \pm Q^2$ (two blue dashed boxes), the sides (four red boxes, solid lines), and the top and bottom edges (black dashed boxes). We then find
\begin{align}
\nonumber G(Q) &= 2 \times Q^2 + 4 \times \left[\sum_{j = 1}^{Q-1} (Q-j/2)^2\right] + 2 \times \left[\sum_{k = 0}^Q (Q/2-k)^2 \right] \\
\nonumber      &= 2 \times Q^2 + 4 \times \left[ \frac{14 Q^3 - 15 Q^2 + Q}{24} \right] + 2 \times \left[ \frac{Q^3 + 3 Q^2 + 2 Q}{12}\right] \\
\nonumber  &= \frac{5}{2} Q^3 + \frac{1}{2} Q
\end{align}
and hence $\chi(Q) = \sum_{q=1}^{Q} G(q) = (5/8) Q^4 + (5/4) Q^3 + (7/8) Q^2 + (1/4) Q$, as we state in the main paper. 

How does $\chi$ scale with the cluster size? A cluster with $Q$ layers has $N(Q) = 1 + 3 Q + 3 Q^2$ cells, so $\chi \sim Q^4 \sim N^2$. For a roughly circular cluster of radius $R$, we'd then expect that $\chi \sim R^4$. This will obviously depend on the precise details of the cluster shape and cell-cell spacing, but if we can approximate the sum in $\chi = \frac{1}{2}\sum_i |\dri|^2$ as an integral, we find $\chi \approx \frac{1}{2}\rho_c \int d^2 r r^2 = \frac{\pi}{4} \rho_c R_{\textrm{cluster}}^{4}$, where $\rho_c$ is the number of cells per unit area in the cluster.  

{How does this scaling compare with earlier results for single cells? Ref. \cite{hu2011geometry} uses a maximum likelihood method to determine the accuracy limit for a single cell with $n_r$ receptors spaced around its radius, finding $\sigma_p^2 \sim \gamma / n_r$, where $\gamma = \frac{(\bar{c}+K_D)^2}{\bar{c}K_D}$ and $p \sim g \times R_\textrm{cell}$, with $R_\textrm{cell}$ the cell radius. Changing variables to $\vb{g}$, this is $\sigma_g^2 \sim \frac{\gamma}{n_r R_\textrm{cell}^2}$. Our results, ignoring cell-to-cell variation, are that $\sg^2 \sim \frac{\gamma}{n_r \chi}$, with $\chi \sim \rho_c R^4$ for a circular cluster of radius $R$, and $n_r$ the number of receptors per cell. Then, as there are $N \sim \rho_c R^2$ cells in the cluster, the total number of receptors within the cluster is $n_t \sim N\times n_r$. Our result for $\sg$ (in the absence of time averaging and CCV) is then $\sg^2 \sim \frac{\gamma}{n_t R^2}$. This shows that, up to geometric factors, a cluster's sensing bound is the same as a giant cell with the same {\it total} number of receptors and radius.} 

\section{Detailed derivation of time-averaged gradient sensing error}
\label{app:timeaverage}

We will, in this section, completely neglect the ligand-receptor fluctuations in gradient sensing, as appropriate for the physically likely case $\sd > 0.1$. We show in Appendix \ref{app:mle} that the maximum likelihood estimator of the gradient is then $\hat{\gv} = (\mathcal{A}^{-1})\cdot \sum_i (M^i - 1) \dri $, where $\mathcal{A}_{\alpha\beta} = \sum_i \delta r_\alpha \delta r_\beta$, with $\alpha,\beta$ the Cartesian coordinates $x,y$. 

We now evaluate $\hat{\gv}$ with the signal $M^i = 1 + \dri\cdot\gv + \Delta^i$:
\begin{align}
\hat{\gv} &= \gv + (\mathcal{A}^{-1})\cdot\sum_i \Delta^i \dri \\
		  &\equiv \gv + \lam 
\end{align}
We will treat a more general case than in the main body, allowing $\Delta_i$ to vary in a time-dependent manner, $\davg{\Delta^i(t) \Delta^j(0)} = \sd^2 C_{\Delta\Delta}(t) \delta^{ij}$ where $\delta^{ij}$ is the Kronecker delta function and $C_{\Delta\Delta}(t)$ characterizes the correlation of the CCV; within the main paper, we take $C_{\Delta\Delta} \to 1$, assuming CCV is persistent over all relevant time scales of the motion. 

Suppose the cell time-averages its maximum likelihood estimator with a time window \tavg:
\begin{align}
\ght(t) &= \int \gh(t') K_T(t-t') \\
        &= \gv + \lamt
\end{align}
where $K_T(t)$ is an averaging function with $K(t<0) = 0$ and $\lamt \equiv \int_{-\infty}^{\infty} \lam(t') K_T(t-t') dt'$. We will often use a simple exponential average with $K(t) = \theta(t) \frac{1}{T} e^{-t/T}$, where $\theta(t)$ is the Heaviside step function. 

Clearly, for a single configuration of cells, $\davg{\ght} = \gv$.  We then would like to compute how much the variations in $\ght$ are reduced by the time-averaging, i.e. we compute

\begin{align}
\davg{ |\ght-\gv|^2 } \equiv \sgt^2 &= \davg{ |\lamt|^2 } \\
									 &= \davg{ \int_{-\infty}^{\infty}dt'\int_{-\infty}^{\infty}dt'' \lam(t')\cdot\lam(t'') K_T(t-t') K_T(t-t'')} \\
									 &=  \int_{-\infty}^{\infty}dt''\int_{-\infty}^{\infty}dt'' K_T(t-t') K_T(t-t'') \davg{\sum_{i,j} \Ainv_{\alpha\gamma}(t') \Delta^i(t') \delta r_\gamma^i(t') \Ainv_{\alpha\beta}(t'') \Delta^j(t'') \delta r_\beta^j(t'')} 
\end{align}
where we have used Einstein summation notation in the last equation. We also note that the average $\davg{\cdots}$ now includes an average over time -- the cell configurations are changing. We now want to perform the average over the $\Delta^i$. This is possible if the re-arrangement of the cell positions are independent of the particular values of $\Delta$, i.e. $\davg{\Ainv_{\alpha\gamma}(t') \Delta^i(t') \delta r_\gamma^i(t') \Ainv_{\alpha\beta}(t'') \Delta^j(t'') \delta r_\beta^j(t'')} \approx \davg{\Delta^i(t')\Delta^j(t'')} \langle \Ainv_{\alpha\gamma}(t') \delta r_\gamma^i(t') \Ainv_{\alpha\beta}(t'') \delta r_\beta^j(t'')  \rangle$.  This would be natural if, e.g. the cluster collectively chooses an estimated direction, but the re-arrangements are only due to local fluctuations, independent of $\Delta$. However, if each cell has a motility related to $\Delta^i$, this assumption may not be accurate. This approximation is also slightly violated if the cell cluster takes on a different shape in response to its estimate of the gradient location. This is an important approximation, but one that we suspect is unavoidable to create a measure of the correlation $\davg{\lam(t) \lam(t')}$ that does not depend on $\Delta$. With this decoupling approximation, we find:
\begin{align}
\sgt^2 &\approx  \int_{-\infty}^{\infty}dt'\int_{-\infty}^{\infty}dt'' K_T(t-t') K_T(t-t'') \sum_{i,j}\davg{\Delta^i(t')\Delta^j(t'')} \langle \Ainv_{\alpha\gamma}(t') \delta r_\gamma^i(t') \Ainv_{\alpha\beta}(t'') \delta r_\beta^j(t'')  \rangle\\
&= \sd^2 \int_{-\infty}^{\infty}dt'\int_{-\infty}^{\infty}dt'' K_T(t-t') K_T(t-t'') C_{\Delta\Delta}(t'-t'') \sum_i \langle \Ainv_{\alpha\gamma}(t') \delta r_\gamma^i(t') \Ainv_{\alpha\beta}(t'') \delta r_\beta^i(t'')  \rangle \label{eq:inprogress}
\end{align}
We emphasize that in the absence of averaging, $K_T(t-t') \to \delta(t-t')$, our result agrees with the results of Appendix \ref{app:mle}. In this case, the right-hand-side of \eq \ref {eq:inprogress} becomes $\sd^2 C_{\Delta\Delta}(0) \sum_i \langle \Ainv_{\alpha\gamma}(t) \delta r_\gamma^i(t) \Ainv_{\alpha\beta}(t) \delta r_\beta^i(t)  \rangle = C_{\Delta\Delta}(0) \sd^2 \langle \Ainv_{\alpha\alpha} \rangle$, as $\sum_i \delta r_\gamma^i \delta r_\beta^i = \mathcal{A}_{\gamma\beta}$. $C_{\Delta\Delta}(0) = 1$ by definition. This suggests we write
\begin{align}
\nonumber \sgt^2 = \sgz^2 &\times \\ \int_{-\infty}^{\infty}dt''&\int_{-\infty}^{\infty}dt'' K_T(t-t') K_T(t-t'') C_{\Lambda\Lambda}(t'-t'') \label{eq:conv}
\end{align}
where $C_{\Lambda\Lambda}(t)$ is the normalized correlation function $C_{\Lambda\Lambda}(t'-t'') = C_{\Delta\Delta}(t-t')\times \sum_i \langle \Ainv_{\alpha\gamma}(t') \delta r_\gamma^i(t') \Ainv_{\alpha\beta}(t'') \delta r_\beta^i(t'')  \rangle / \langle \Ainv_{\mu\mu} \rangle$.

The correlation function $C_{\Lambda\Lambda}$ can be calculated readily from the cell trajectories relative to the cluster center of mass $\delta \rb^i(t)$. However, in the limit of isotropic clusters of roughly constant shape, we can significantly simplify this form. For isotropic clusters of constant shape, $\mathcal{A}_{\alpha\beta}(t) = \chi \delta_{\alpha\beta}$ independent of time. Given this assumption, $C_{\Lambda\Lambda}(t'-t'') / C_{\Delta\Delta}(t'-t'')  = \chi^{-1} \sum_i \langle \delta r_\alpha^i(t') \delta r_\alpha^i(t'')  \rangle$ -- i.e. $C_{\Lambda\Lambda}(t'-t'') = C_{\Delta\Delta}(t'-t'')C_{rr}(t'-t'')$, where $C_{rr}(t'-t'') \equiv \langle \boldsymbol\delta\rb(t')\cdot\boldsymbol\delta\rb(t'') \rangle / \langle |\boldsymbol\delta\rb|^2 \rangle$. 

The double convolution in \eq \ref{eq:conv} is simpler in Fourier space, 
\begin{equation}
\sgt^2 = \sgz^2 \times \int_{-\infty}^\infty \frac{d\omega}{2\pi} |K_T(\omega)|^2 C_{\Lambda\Lambda}(\omega) \label{eq:gtvar}
\end{equation}
where $K_T(\omega)$ is the Fourier transform of $K_T(t)$, $K_T(t) = \int \frac{d\omega}{2\pi} e^{i\omega t} K_T(\omega)$ and $C_{\Lambda\Lambda}(\omega)$ the Fourier transform of $C_{\Lambda\Lambda}(t)$.  For $K_T(t-t') = \theta(t-t') \frac{1}{T} e^{-(t-t')/T}$, $K_T(\omega) = \frac{1}{1+i\omega T}$. 

In the common case that $C_{\Lambda\Lambda}(t) = \exp(-t/\tau_\Lambda)$, where $\tau_\Lambda$ is a characteristic correlation time, and $K_T(t-t') = \theta(t-t') \frac{1}{T} e^{-(t-t')/T}$, this is even simpler:
\begin{equation}
\sgt^2 = \frac{\sgz^2}{1 + T/\tau_\Lambda} 
\end{equation}
When we can approximate $C_{\Lambda\Lambda} \approx C_{rr}$, this is consistent with our intuition: it takes roughly a time of $\tau_r$ for the cells to re-arrange, and so in a time of $T$, the cluster can make $T/\tau_r$ independent measurements in an averaging time, and so it can decrease the measurement error by $T/\tau_r$. If the amount of CCV varies over time, the characteristic timescale is then $\tau_c = \frac{\tau_r \tau_\Delta}{\tau_r + \tau_\Delta}$ -- the relaxation timescale that is relevant is the faster of the two timescales. Within the main article, we have assumed that this is always cell position rearrangement. 

Because of the complexity of the different results and regimes in this section, we provide a summary in Table \ref{tab:summary}.

\renewcommand{\arraystretch}{2}
\begin{table*}
\begin{tabular}{|l|p{5.6cm}|p{6.4cm}|}
\hline
{\bf Formula for $\sgt^2$}& {\bf Assumptions made} & {\bf Associated quantities} \\
\hline
$\sgz^2 \times \int_{-\infty}^\infty \frac{d\omega}{2\pi} |K_T(\omega)|^2 C_{\Lambda\Lambda}(\omega)$ & Decoupling of averages over CCV and configurations & \parbox[t]{5cm}{$C_{\Lambda\Lambda}(t'-t'') = C_{\Delta\Delta}(t-t')\times$ \\ $\sum_i \langle \Ainv_{\alpha\gamma}(t') \delta r_\gamma^i(t') \Ainv_{\alpha\beta}(t'') \delta r_\beta^i(t'')  \rangle / \langle \Ainv_{\mu\mu} \rangle$} \\
\hline 
$\sgz^2 \times \int_{-\infty}^\infty \frac{d\omega}{2\pi} |K_T(\omega)|^2 \left\{C_{\Delta\Delta}(t) C_{rr}(t)\right\}_\omega$ & Decoupling of averages over CCV and configurations; isotropic clusters of roughly constant shape & $C_{rr}(t'-t'') \equiv \langle \boldsymbol\delta\rb(t')\cdot\boldsymbol\delta\rb(t'') \rangle / \langle |\boldsymbol\delta\rb|^2 \rangle$, $\{f(t)\}_\omega$ is the Fourier transform of $f(t)$ to $\omega$.\\
$\sgz^2/\left[1+T/\tau_\Lambda\right]$ & Decoupling of averages over CCV and configurations, exponential time averaging, $C_{\Lambda\Lambda} = e^{-t/\tau_\Lambda}$ & \\
\hline 
$\sgz^2/\left[1+T/\tau_{c}\right]$ & Decoupling of averages over CCV and configurations, exponential time averaging, isotropic clusters of roughly constant shape, $C_{\Delta\Delta} = e^{-t/\tau_\Delta}$, $C_{rr} = e^{-t/\tau_r}$& $\tau_c = \frac{\tau_r \tau_\Delta}{\tau_r + \tau_\Delta}$ \\
\hline
\end{tabular}
\caption{Summary of different bounds and assumptions made} \label{tab:summary}
\end{table*}

\section{Characteristic time scales of different re-arrangement mechanisms}
\label{app:mechanisms}

The characteristic time of the position-position correlation, $\tau_r$ is critical in calculating chemotactic accuracy via \eq \ref{eq:snrt_simp}. What is $\tau_r$, and what does it depend on? This depends on the mechanism of re-arrangement. 

{\bf Persistent cluster rotation.} If the cluster rigidly rotates with angular velocity $\Omega$, we can see that $\rb(t)\cdot\rb(0) = |\rb(0)|^2 \cos \Omega t$, and time-averaging over a timescale $T > \Omega^{-1}$ can improve the signal-to-noise ratio. How does $\Omega$ depend on the cluster and cell properties? As, for a cluster to actively rotate with velocity $\Omega$, cells at the edge must be able to crawl with speed $R \Omega$, where $R$ is the cluster radius, we expect that $\Omega \sim v_{\textrm{max}}/R$, where $v_{\textrm{max}}$ is the maximum speed a cell can crawl.  Hence, the characteristic time over which $\rb$ changes is $\tau_{\textrm{rot}} \sim R/v_{\textrm{max}}$, as studied explicitly for rotating clusters in the main paper. 

{\bf Rotational diffusion of the cluster.} If the cluster diffuses rotationally as a rigid body with angular diffusion rate $D_r$, $\langle \rb(t)\cdot\rb(0)\rangle = \langle |\rb(0)|^2\rangle e^{-D_r t}$. However, the scaling of the rotational diffusion coefficient with cluster size is not obvious, and will depend on the model.  Rotational diffusion could arise from clusters that undergo collectively-driven rotation, but occasionally switch between moving in different directions over a timescale $\tau_{\textrm{switch}}$ -- this is observed in small clusters of cells on micropatterns \cite{segerer2015emergence}.  If this is the case, the effective rotational diffusion coefficient is just $D_r \sim \Omega^2 \tau_{\textrm{switch}}$ -- so $D_r \sim v_{\textrm{max}}^2 \tau_{\textrm{switch}}/R^2$.  By contrast, we have seen that in the absence of a collective aligning effect, rotational diffusion can be small or absent, depending on certain details about the underlying cell motility model \cite{camley2016emergent}.

{\bf Cell re-arrangements.} Cells within a tissue can often be described as undergoing a persistent random walk: they maintain a direction over a timescale $\tau_{\textrm{persist}}$, but lose their orientation beyond this time, leading to an effectively diffusive motion with effective diffusion coefficient $D_{\textrm{eff}}$ at long time scales \cite{szabo2010collective}; these numbers can be on the order of 10 microns$^2$/min.  For cells to move from one half of the cluster to another by diffusion will then take a timescale $\tau_\textrm{diff}$ where $D_{\textrm{eff}} \tau_{\textrm{diff}} \sim R^2$.  However, if cells are persistent over the timescale required to cross the cluster, re-arrangements could be accelerated -- in this case, $\tau_{\textrm{rearrange}} \sim R/v_\textrm{cell}$. Naturally, if cell motion becomes subdiffusive, this will also change the dynamics of the cluster re-arrangement.  This would all be captured in \eq \ref{eq:snrt} once the cell-cell correlation function is determined.

\section{Bounds capture variation of CI over large range of parameters}
\label{app:params}

 \begin{figure*}[ht!]
 \centering
 \includegraphics[width=180mm]{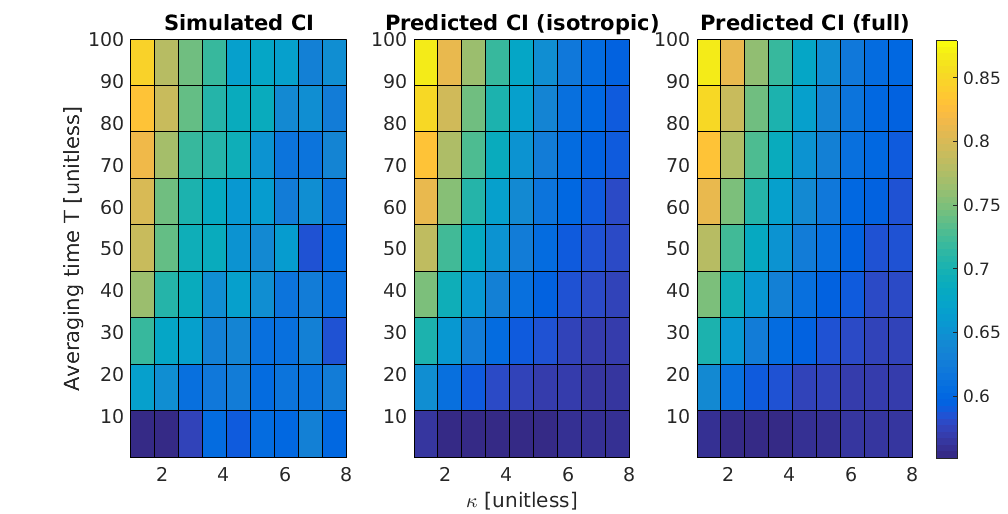}
  \includegraphics[width=180mm]{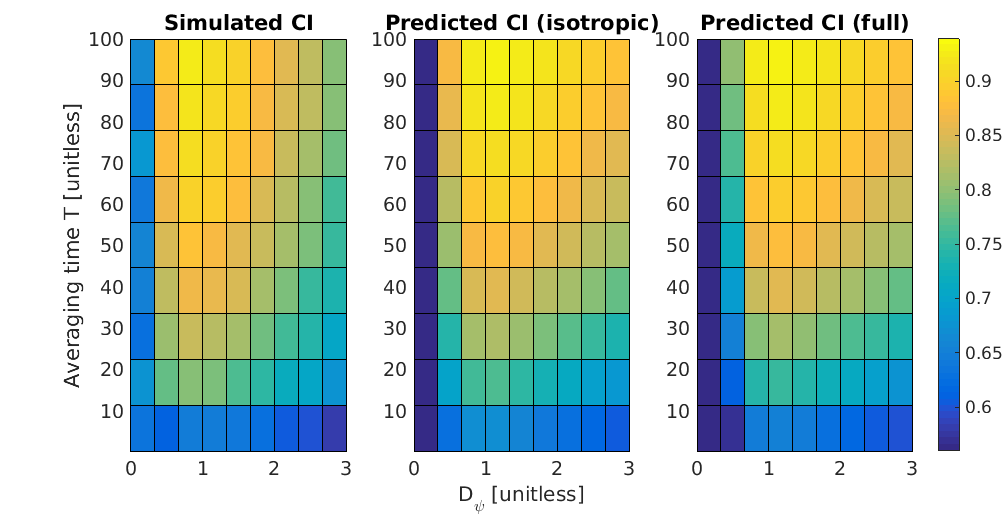}
 \caption{\linespread{1.0}\selectfont{} {\bf Variation of bound captures simulated CI well}. Parameters not varied match those used in \fig \ref{fig:Dpsi}, as does the simulation setup, but with $D_\psi = 1$ (TOP) and $\kappa = 0.25$ (BOTTOM).}
 \label{fig:compareall}
 \end{figure*}

In this section, we show a larger range of variations in parameters, showing that the predicted CI from the bound captures simulated CI in our models well. 
While our primary results on computing the upper bound in Section \ref{sec:fluidity} were performed with \eq \ref{eq:snrt_simp}, it is also possible to compute $\sg^2$ without any assumptions about cluster isotropy, using the correlation function $C_{\Lambda\Lambda}(t)$ as seen in Appendix \ref{app:timeaverage}, and the no-time-averaging result of $\sigma_{\mathbf{g},0}^2$ via \eq \ref{eq:traceerr}. We find that typically, in our simple collective cell motility simulations, that $\tau_{\Lambda}$ and $\tau_r$ are very close, as are $\textrm{tr} \mathcal{A}^{-1}$ and $2/\chi$ -- unless the cluster becomes elongated or otherwise anisotropic. However, we do note that $\textrm{tr} \mathcal{A}^{-1} \ge 2/\chi$ for any configuration of cells, so this implies a larger uncertainty than the isotropic approximation. We compare the simulated CI and our predictions from the isotropic and full theories in \fig \ref{fig:compareall}.

\end{document}